\documentclass{aa}  

\usepackage{graphicx}
\usepackage{txfonts}
\usepackage{amsmath}
\usepackage{siunitx}
\usepackage{xcolor}
\usepackage{pgfplots} 
\usepackage{longtable} 
\usepackage[online]{threeparttablex} 
\usepackage{array} 

\usepackage{hyperref}

\newcommand\inputpgf[2]{{
\let\includegraphicsWithoutPath\includegraphics
\renewcommand{\includegraphics}[2][]{\includegraphicsWithoutPath[##1]{#1/##2}}
\input{#1/#2}
}}
\pgfplotsset{compat=1.17} 

\usepackage{adjustbox} 
\usepackage{printlen} 
\usepackage{hyperref,xcolor} 
\usepackage{supertabular} 
\usepackage{booktabs} 

\newcommand{\comment}[1]{}

\begin{document}

   \title{Gaia search for early-formed andesitic asteroidal crusts} 

   \author{M. Galinier\inst{1}
          \and
          M.~Delbo \inst{1}
          \and
          C. Avdellidou\inst{1}
          \and
          L. Galluccio\inst{1}
          \and
          Y. Marrocchi \inst{2}}

\institute{Universit\'e C\^ote d'Azur, CNRS--Lagrange, Observatoire de la C\^ote d'Azur, CS 34229 -- F 06304 NICE Cedex 4, France\\
\email{marjorie.galinier@oca.eu}
\and
CRPG, CNRS, Université de Lorraine, UMR 7358, Vandoeuvre-les-Nancy F-54501, France
             }

   \date{Received date, year; accepted date, year}

 
  \abstract
   {Andesitic meteorites are among the oldest achondrites known to date. They record volcanic events and crust formation episodes in primordial planetesimals that took place about 4.565~Myr ago. However, no analogue for these meteorites has been found in the asteroid population to date.}
   {We searched for spectroscopic analogues of the andesitic meteorite Erg Chech 002 in the asteroid population using the Gaia DR3 spectral dataset.}
   {In order to identify which asteroids have the most similar spectrum to Erg Chech 002, we first determined the spectral parameters of Gaia DR3 asteroids (spectral slope and Band I depth) and compared them to the spectral parameters of different samples of the meteorite. In addition, we performed a spectral curve matching between Erg Chech 002 and Gaia DR3 asteroid data, and we compared the results of both methods.}
   {We found that 51 main-belt asteroids have a visible spectrum similar to the one of Erg Chech 002, and 91 have a spectrum similar to the space-weathered spectra of the meteorite, corresponding to 0.08 and 0.15\% of the whole Gaia DR3 dataset of asteroids with spectra, respectively. The asteroids that best match the laboratory samples of the meteorite are mostly located in the inner main belt, while the objects matching the space-weathered meteorite models show slightly more scattering across the belt.}
   {Despite the fact that we find asteroids that potentially match Erg Chech 002, these asteroids are extremely rare. Moreover, a visible spectrum alone is not completely diagnostic of an Erg Chech 002-like composition. Near-infrared spectra will be important to confirm (or rule out) the spectral matches between Erg Chech 002 and the candidate asteroid population.}


   \maketitle
%

\section{Introduction}


Planetesimal accretion is considered the first stage of planetary formation. The composition and sizes of these planetesimals and the heliocentric distance of their accretion are key long-standing issues in planetary science \citep[see, e.g.][and references therein]{johansen2015}.
Planetesimal accretion took place during the first million years of our Solar System's history \citep{henke2012,trieloff2022, morbidelli2020, morbidelli2022}, and the planetesimals that formed at the earliest times are expected to have been highly heated by the radioactive decay of $^{26}$Al, and thus to be differentiated. During this process, the interior of a molten body with an initial homogeneous composition organises into layers of different densities and compositions, forming a dense metallic core, an olivine-rich overlaying mantle, and an igneous crust \citep[e.g.][and references therein]{mcsween2002}. Subsequent collisional evolution fragmented those original planetesimals, producing families of asteroid fragments. These fragments should show different physical and spectral properties depending on the type of collision and the depth of the material excavation during the impact event. Moreover, family fragments can drift towards regions of orbital instability due to non-gravitational forces and then be delivered to Earth as meteorites. 

Meteorites show a large range of compositions, reflecting the composition of the different layers of the parent body from which they are derived, if differentiated \cite[e.g.][]{greenwood2020}. Linking meteorites to asteroids provides insights into the internal structure of the parent body and into its accretion time and region. However, only a few links have been established up to now: the Howardite-Eucrite-Diogenite meteorites (HEDs) have been linked to the asteroid (4) Vesta and its family \citep{russell2012}; the aubrite meteorites (enstatite achondrites) have been connected to the (434) Hungaria family \citep{lucas2019}; and very recently the enstatite chondrite meteorites of EL type were linked to the asteroid family of (161) Athor \citep{avdellidou2022}. All of these families are located in the inner main belt (i.e. with a semi-major axis between 2.1 and 2.5 au).

The study of HEDs, for example, showed that they originate from the igneous crust of asteroid (4) Vesta \citep[e.g.][]{McCord1970,russell2012,BinzelXu1993,Burbine2001,desanctis2012,russell2013}, which is known to be differentiated \citep[e.g.][]{Ruzicka1997,Righter1997,MandlerElkins-Tanton2013}. However, other eucrite meteorites that do not belong to the HEDs and thus do not come from Vesta have also been identified \citep{Bland2009}. Moreover, lithological, colour, and albedo differences have been detected by \cite{Mansour2020} between the vestoids, other low inclination basaltic asteroids of the inner belt, as well as basaltic asteroids with orbits beyond 2.5 au. All of these point to the necessary existence of another basaltic source of meteorites. \cite{Oszkiewicz2015} suggest that this object could be the parent body of the Flora family.

Despite the evidence given by the meteorites, few signs of differentiation amongst asteroids have been found to date. Searches for a population of basaltic crust-like asteroids \citep[in and outside the Vesta family; e.g.][]{moskovitz2008,solontoi2012,leith2017} as well as metallic ones \citep[e.g.][]{harris2014} have been successful. However, there is an observational lack of mantle-like olivine-rich asteroids in the main belt \citep{DeMeo2019}. These asteroids are identified as A types \citep{bus2002tax,demeo2009}; in addition compared to the amount of basaltic and metallic asteroids in the main belt, they should be found in a larger proportion than what has been observed so far. This long-standing issue in planetary science is the so-called missing mantle problem \citep{chapman1986}.

Another interesting class of meteorites has been recently identified as evidence of differentiation in the main belt, in addition to the aubrites, iron meteorites, HEDs, and eucrites: the so-called andesitic meteorites \citep{Day2009, barrat2021}. The formation mechanism of these meteorites is consistent with rapid cooling of a silicate-rich magma at the surface of a planetesimal. However, said mechanism is still debated \citep{Arculus2009}. In particular, the meteorite Erg Chech 002 (hereafter EC~002) found in May 2020 in the Sahara desert is reported by \cite{barrat2021} to be 'the oldest andesite of the Solar System', with a measured crystallisation age of 4,565~Myr (around 2.25~Myr after the beginning of the Solar System). It has been classified as an ungrouped achondrite and it is spectroscopically unique. Its composition is similar to those experimentally produced by low partial melting of ordinary chondrite-like materials \citep{CollinetGrove2020}. This suggests that EC~002 could originate from the igneous crust of a non-carbonaceous planetesimal that suffered from low partial melting. This crust is thought to have been separated from the original parent body by a violent event, as suggested by evidence of a rapid cooling. EC~002 was thrown into space and travelled as part of a bigger body, before separating from it. As its composition is different from the HEDs, this meteorite provides evidence that some planetesimals were covered in andesitic and not basaltic crusts, the process of differentiation thus being different for these bodies.

The parent bodies of andesitic meteorites and planetesimals with andesitic crusts are unknown to date. \cite{barrat2021} searched for objects with similar properties to EC~002 among the main belt asteroid population. To do so, they compared laboratory spectra of different samples of EC~002 to astronomical spectra of asteroids with strong pyroxene signatures, namely taxonomic end members of classes O and V, and to spectrophotometric data from the Sloan Digital Sky Survey (SDSS). No satisfying match between the meteorite and the asteroids was found. The authors concluded that almost the entire original population of planetesimals must have disappeared, as well as their fragments. They speculate that the disappearance of EC~002-like objects could be due either to their accretion to other asteroids to form larger planetary embryos, or to their destruction. This could also result from their erasure by subsequent stages of melting and planetary accretion and differentiation \citep{CollinetGrove2020}. \\

Reflectance spectra of EC~002 were acquired by \cite{barrat2021}. These spectra show the presence of two strong absorption bands that were linked to Ca-rich pyroxene: a first band centred around 0.95~\SI{}{\micro\meter} (Band I), and a second one around 2~\SI{}{\micro\meter} (Band II). They also show the presence of a small band centred around 0.65~\SI{}{\micro\meter}, whose origin is not discussed by \cite{barrat2021}. However, the analysis done by the authors show that the pyroxenes of EC~002 are quite rich in Cr-bearing species (as shown in Table~S2 of their supplementary material); and according to \cite{moskovitz2008}, \cite{cloutis2018} and \cite{cloutis2002}, Cr-rich high-Ca pyroxene can lead to the apparition of absorption features near 450 and 0.65~\SI{}{\micro\meter}. Thus, the 0.65~\SI{}{\micro\meter} absorption band observable in the reflectance spectrum of EC~002 could be due to the presence of chromium in the pyroxene of the meteorite. Comparing this spectrum with available asteroids spectra, the authors found no known asteroid spectral type presenting such absorption signatures.

In this work we take advantage of the recent publication of an unprecedented sample of asteroid spectra by the Data Release 3 (DR3) of the ESA mission Gaia \citep{galluccio2022} to search for analogues of EC~002 in the asteroid population. In Sect.~\ref{data} the dataset of asteroid spectra used is presented, along with the spectral data of the meteorite retrieved from \cite{barrat2021}. The methods are detailed in Sect.~\ref{methods}. In Sect.~\ref{results} we present our results, followed by a discussion in Sect.~\ref{discussion}.


\section{Data}
\label{data}

In order to search for analogues of EC~002 among the asteroid population, we used the dataset of reflectance spectra that was acquired by Gaia between 5 August 2014 and 28 May 2017, and was released in June 2022. This dataset consists of mean reflectance spectra in the visible wavelength range of 60\,518 Solar System objects (SSOs), with the majority of objects having magnitudes between $\simeq$18 and 20. This is an unprecedented dataset of objects that are usually too faint to be observed from ground-based telescopes.

The reflectance spectra are acquired by two low-resolution slit-less spectrophotometers on board Gaia, the blue and red spectrophotometers (BP and RP), which are respectively optimised for the blue and red part of the spectrum. Specifically, the BP spans the wavelength range from 0.33 to 0.68~\SI{}{\micro\meter} and the RP covers the range from 0.64 to 1.050~\SI{}{\micro\meter}. The spectral resolution of each spectrophotometer is a function of wavelength, and varies from 4 to 32~nm~pixel$^{-1}$ for the BP and 7 to 15~nm~pixel$^{-1}$ for the RP \citep{galluccio2022,carrasco2021,jordi2010}. When an SSO transits on the focal plane of Gaia at a given epoch, each spectrophotometer measures counts at every wavelength to create `epoch spectra'. Given that the wavelength range of both instruments overlaps in the 0.65-0.68~\SI{}{\micro\meter} interval, the two epoch spectra are merged to create a full epoch spectrum. To each SSO is associated a unique mean reflectance spectrum obtained by averaging several epoch spectra, spanning the visible wavelength range from 374 to 1034~nm in 16 discrete wavelength bands \citep{galluccio2022}. A `spectral\_validation\_flag' (hereafter flag) number is associated to each band, assessing the estimated quality of the band. In some cases, the merging of the epoch spectra taken by each spectrophotometer is not perfect and can lead to the creation of artefact bands \citep[see][]{galluccio2022}. Caution must thus be taken when analysing the mean reflectance spectra in the overlapping wavelength interval. In a similar way, the bluest and reddest data bands of Gaia spectra are in general affected by large systematics due to the low efficiency of the spectrophotometers in these bands. They are not always flagged but they need to be taken with caution as well.
To assess the quality of the asteroid analogues of EC~002 found using Gaia data, visible (VIS) and near-infrared (NIR) spectra from the literature were used in our analysis. 

To perform our analysis, we used the EC~002 spectra that were published by \cite{barrat2021}. \cite{barrat2021} acquired visible and near-infrared reflectance spectra of one powder sample and three raw slabs samples of EC~002. The spectra were digitised from the supplementary material of \cite{barrat2021} using the region features (points and box) of the SAO Image DS9 software. A python code was used to transform the pixel coordinates to reflectance and wavelengths units, and we verified that our digitised spectra were indistinguishable from the original ones before conducting the study. The spectra of the laboratory samples of the meteorite were later kindly provided to us by Jean-Alix Barrat and his co-authors (J. A. Barrat, P. Beck, private communication).
Since asteroid surfaces can be altered by space weathering and in order to compare the meteorite spectrum with asteroid spectra, \cite{barrat2021} applied a space weathering model \citep{Hapke2001} to the powder sample of the meteorite, to simulate the effects of solar wind ion bombardment and micrometeorite impacts on the surface of the body. They published three space-weathered spectra of EC~002 corresponding to three different levels of space weathering -- low, medium, and high -- which we retrieved using SAO image DS9 software. In our study, we used the four reflectance spectra of the laboratory samples of the meteorite and the three modelled space-weathered spectra to search for asteroids with similar features to EC~002.

In order to compare our work with the one of \cite{barrat2021}, we also performed some comparison tests between the SDSS and the Gaia dataset. The SDSS dataset used in this work contains information for 33\,584 asteroids and was retrieved from the work of \cite{demeo2013}. No selection criteria was applied to filter out noisy data, regardless of the uncertainties of the SDSS dataset.


\section{Methods}
\label{methods}

In order to identify a spectral link between EC~002 and Gaia asteroids, we first compared the laboratory spectra of EC~002 to Gaia spectra without considering the effect of space weathering. Since the source of this meteorite is yet unknown, there is a probability that this object originates from a family of young fragments created by a recent collision. 
These fragments would have suffered limited space weathering because of their young age, showing a spectrum similar to the one of EC~002. Moreover, we present here an attempt to detect asteroids with similar spectral features as of EC~002 (similar spectral slope, presence of a pyroxene band around 0.95~\SI{}{\micro\meter} and of a small 0.65~\SI{}{\micro\meter} band), and these features are more easily detected without space weathering. It is also reasonable to believe that asteroids have surface grains. Given that they influence the spectroscopic properties of a medium, we studied the spectra of the powder and raw slab samples of EC~002.

On the other hand, EC~002 has the composition of a partial melt of an ordinary chondrite \citep{barrat2021}. Once weathered, ordinary chondrites are spectrally similar to S-type asteroids. If the asteroids matching EC~002 suffered from space weathering, it is not unreasonable to expect a S-type-like space weathering \citep[as expressed by the space weathering trend assumed by][]{barrat2021}. Hence, we studied in a second time the modelled space-weathered spectra of EC~002. To summarise, we searched for asteroids spectrally matching the powder, raw slabs samples, and modelled space-weathered spectra of EC~002. To do so, we used two spectral matching methods described below. 

\subsection{Band I depth vs. slope comparison}

The first method consists in comparing the spectral parameters derived from the reflectance spectra of the meteorite and of the asteroids. These parameters are the slope of the reflectance spectrum between 468.6 and 748~nm, and a measure of the depth of the silicate band centred around 950~nm (Band I depth). This method was inspired by the works of \cite{barrat2021}, \cite{demeo2013} and \cite{parker2008} using the SDSS asteroid spectrophotometric data. \cite{ivezic2001} and \cite{Nesvorny2005} performed a principal component analysis on these data and identified two spectral parameters that express most of the data variability: the a* parameter and the i-z colour. The a* parameter closely represents the slope of the reflectance spectrum in the g', r' and i' SDSS bands \citep{ivezic2001}, these bands being respectively centred at 468.6~nm, 616.6~nm and 748.0~nm \citep{demeo2013}. The i-z colour is sensitive to the depth of a potential 1~\SI{}{\micro\meter} band, the colour being the difference of magnitude between the i' and z' bands. These parameters are useful to characterise a visible asteroid spectrum.

\cite{demeo2013} used slightly different spectral parameters to characterise the asteroids: the z-i parameter and the gri-slope. To evaluate them, the SDSS observed magnitudes of asteroids are converted into reflectance values at the centre of each SDSS filter, and the derived reflectance spectra are normalised to unity at the central wavelength of the g' filter (468.6~nm). The gri-slope is defined as the slope of the derived reflectance spectra over the g', r' and i' filters. The z-i parameter still measures the depth of a potential 1~\SI{}{\micro\meter} band, but it is here defined as: 

\begin{equation}
\label{eq:RzRi}
    \mathrm{R_z-R_i=R(\lambda = 893.2~nm)-R(\lambda = 748.0~nm)}.
\end{equation}

The gri-slope and z-i colour of asteroids have been used to group objects into classes since \citep{Ivezic2002,carvano2010,demeo2013}. We note that the parameter measuring the Band I depth used in this study is a difference of reflectance, we therefore refer to it as $\mathrm{R_{z}-R_{i}}$.

In order to compare Gaia reflectance spectra with what has been done in the work of \cite{barrat2021}, we evaluated the gri-slope and $\mathrm{R_{z}-R_{i}}$ parameters for Gaia spectra. First, each Gaia spectrum was interpolated using a cubic smoothing spline (python3 package csaps, default smoothing parameter) and re-sampled between 450 and 900~nm. During this smoothing procedure, only Gaia bands with flags equal to zero (good quality bands) were considered and the first and last Gaia bands were not taken into account, in order to limit the impact of low quality bands on the calculated reflectance values. The re-sampled Gaia spectra were then normalised at 468.6~nm, and the spectral gri-slope was computed by linearly fitting the spectrum between 468.6 and 748.0~nm (first-degree polynomial fit, numpy package polyfit). The $\mathrm{R_{z}-R_{i}}$ parameter was computed by taking the value of the reflectance of every re-sampled normalised Gaia spectra at the central wavelength of the i' and z' SDSS filters, namely 748.0~nm and 893.2~nm.
The asteroids defined as potentially matching the spectrum of EC~002 are those in an area close to the meteorite in the $\mathrm{R_{z}-R_{i}}$ vs. gri slope diagram. This will be further discussed in section~\ref{results}.

\subsubsection{Asteroid matching with spectra of laboratory samples of EC~002}

First, we studied the four laboratory samples of the meteorite EC~002 without taking space weathering into account. To compare the meteorite spectra with those of Gaia asteroids, their spectral slope and $\mathrm{R_{z}-R_{i}}$ parameter were calculated. To do so, the spectrum of each sample was interpolated between 450 and 900~nm using a cubic smoothing spline (python3 package csaps, default smoothing parameter), as was done for the Gaia data. The spectra were then normalised to unity at 468.6~nm and the $\mathrm{R_{z}-R_{i}}$ parameter was calculated using Eq.~\ref{eq:RzRi}. The spectral slope was evaluated between 468.6 and 748.0~nm applying a first degree polynomial fit.

In order to identify the asteroids with spectral parameters similar to EC~002, we calculated the average of the slope and $\mathrm{R_{z}-R_{i}}$ values for the four samples. The corresponding point is considered as the 'barycentre' of the non-space-weathered samples. Then, we determined a 3$\sigma$ confidence ellipse around this barycentre. The equation of the confidence ellipse centred on a barycentre of coordinates ($\mathrm{x_c}$, $\mathrm{y_c}$) and oriented with an angle $\alpha$ is:
\begin{align} 
\mathrm{ \left( \frac{\cos^2{\alpha}}{a^2} + \frac{\sin^2{\alpha}}{b^2} \right)(x-x_c)^2 + \left( \frac{\sin^2{\alpha}}{a^2} + \frac{\cos^2{\alpha}}{b^2} \right)(y-y_c)^2 + \nonumber}\\ 
\mathrm{2(x-x_c)(y-y_c)\sin{\alpha}\cos{\alpha} \left( \frac{1}{b^2} - \frac{1}{a^2} \right) = \mathrm{s}} ,
\end{align}
with $x$ the gri-slope of a reflectance spectrum, $y = \mathrm{R_{z}-R_{i}}$, and $s$ the scale of the ellipse that represents a chosen confidence level. a and b are respectively the semi-major and semi-minor axis of the ellipse. It is possible using $\chi$-square probabilities to determine that for a 3$\sigma$ ellipse, the $s$ value is 9.210 (99\% confidence level). If an asteroid falls inside the 3$\sigma$ ellipse in the spectral parameter space, then it would be considered as a candidate match of EC~002 according to its visible spectrum.

To compute the parameters of the 3$\sigma$ confidence ellipse, we calculated the covariance matrix of the four laboratory samples of the meteorite. The semi-major and semi-minor axis of the ellipse are defined as:
\begin{equation}
\begin{cases}
\mathrm{a}  &= \mathrm{\sqrt{s \lambda_{1}}} \\
\mathrm{b} &= \mathrm{\sqrt{s \lambda_{2}}},
\end{cases}
\end{equation}
with $\mathrm{\lambda_{1}}$ and $\mathrm{\lambda_{2}}$ the eigenvalues of the covariance matrix, $\mathrm{\lambda_{1}}$ being the largest. The angle $\alpha$ of the ellipse is defined as $\mathrm{\alpha = \arctan\frac{v_1(y)}{v_1(x)}}$ with $\mathrm{v_{1}}$ the eigenvector of the covariance matrix associated to the largest eigenvalue.

\subsubsection{Asteroid matching with modelled space-weathered spectra of EC~002}

After studying non-space-weathered samples of EC~002, we analysed the modelled spectra of EC~002 from \cite{barrat2021} on which was applied the \cite{Hapke2001} space weathering model \citep[from Fig.~S12 of the supplementary material of][]{barrat2021}.
The slope and $\mathrm{R_{z}-R_{i}}$ parameters were calculated for these spectra, following the same procedure as explained before.
In order to study more stages of space weathering of the meteorite, we fitted a straight line to the points corresponding to the powder sample and to the three space-weathered samples in the $\mathrm{R_{z}-R_{i}}$ vs. slope plot. This line will be referred in the following as the `space weathering line'.
A parallel line to the space weathering line centred on the barycentre of the non-weathered samples was calculated, and the 3$\sigma$ ellipse was moved along this line from the lowest to the highest space weathering points, in order to define a 'possible matches area' within which objects could present spectral parameters similar to those of EC~002 with different levels of space weathering. The spectra of the asteroids within this `possible matches area' were then visually inspected. Indeed, we preferred relying on visual inspection rather than on an automated method to assess the quality of the matches, firstly because of the relatively small number of objects to inspect and secondly because the 0.65~\SI{}{\micro\meter} band present on the meteorite spectrum was never detected by algorithms. This band being a characteristic feature of the meteorite spectrum, we chose the method where its presence was the most surely detected.

\subsection{Curve matching}

The second method we used in order to find spectral analogues of EC~002 is a curve matching method, between the meteorite and the asteroids reflectance spectra. This method is widely used in the literature \citep[see, e.g.][]{popescu2012,demeo2022}. It consists in evaluating how similar two spectra are relying on the measure of a best-fit coefficient. In this work, among the possible existing coefficients, we chose to use the $\chi^2$ goodness-of-fit test. 
The reduced $\chi^2$ we used is expressed as:
\begin{equation}
    \mathrm{\chi_{red}^{2} = \frac{1}{\nu}  \sum_{i}^{N} \frac{(A_{i}-f . M_{i})^{2}}{\sigma_{i}^2}},
    \label{eq:chi2_f}
\end{equation}
with $M_{i}$ the meteorite spectrum, $A_{i}$ a Gaia asteroid reflectance spectrum and $\sigma_{i}$ its associated uncertainties, $\nu$ the number of degrees of freedom, and $f$ a normalisation factor allowing the best overlap between the meteorite and Gaia reflectance spectra. 
The $f$-value was determined by minimising the $\mathrm{\chi_{red}^2}$ such that the partial derivative of the $\mathrm{\chi_{red}^2}$ with respect to $f$ is zero. This leads to:
\begin{equation}
    \mathrm{f = \frac{\sum_{i}^{N} \frac{M_{i}.A_{i}}{\sigma_{i}^2}}{\sum_{i}^{N} {\frac{M_{i}^2}{\sigma_{i}^2}}}}.
    \label{formula:f}
\end{equation}

In order to compare EC~002 with asteroids, we started by sampling the meteorite spectrum at Gaia's wavelengths. We considered only the good quality bands in Gaia spectra. Since the bands at the extreme wavelengths of the spectral range are often damaged due to the low BP-RP sensitivity there, the first and last bands of Gaia spectra were not taken into account. Moreover, as explained earlier, DR3 SSO spectra were assigned a non-zero flag to the bands where some potential problems were detected. Every band flagged with a non-zero number was removed. The `cleaned' Gaia spectra were thus composed of 14 bands spanning the wavelength range from 418 to 990~nm, provided that they had a flag=0.

\begin{figure}[h!]
  \centering
  \begin{adjustbox}{clip,trim=0.1cm 0cm 1cm 0.7cm,max width=\textwidth}
  \input{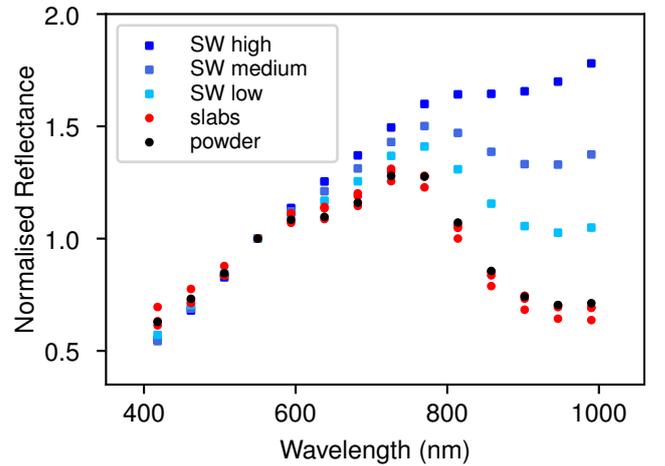}
  \end{adjustbox}
    \caption{Spectra of a powder, three raw slab samples and three modelled spectra of space-weathered of EC~002 sampled and normalised as Gaia data, after removing the first and last Gaia bands. }
\label{fig:EC002_sampled_Gaia}
\end{figure}

Then, a cubic smoothing spline was applied to the meteorite spectrum to interpolate it (python package csaps, smoothing parameter of 0.0001) and the interpolated spectrum was sampled as each cleaned Gaia spectrum. This re-sampled meteorite spectrum was then normalised at 550~nm. Figure \ref{fig:EC002_sampled_Gaia} shows the different spectra of EC~002 normalised and re-sampled.

For each sample of the meteorite, the $\chi_{red}^2$ of Eq.\ref{eq:chi2_f} was calculated between each cleaned asteroid spectrum and the re-sampled and re-normalised meteorite spectrum.
As explained in previous studies \citep{Hanus2015,Hanus2018}, a 3$\sigma$ match to the meteorite is defined as an asteroid respecting the following condition: $\mathrm{\chi_{red}^2 < 1 + 3 \sigma}$ with $\mathrm{\sigma = \frac{\sqrt{2\nu}}{\nu}}$ and $\nu$ the number of degrees of freedom.
%
%
For $\nu$=16, $\chi_{red}^2< $ 2.06 $\simeq \chi_{red}^2< $ 2. For each meteorite sample, the best matches were selected according to this criterion and their spectra were then visually inspected.


\section{Results}
\label{results}

In this section we describe the results from (i) the comparison of the spectral slope and Band I depth, and (ii) the curve matching method, between the spectra of EC~002 and that of Gaia asteroids. As it will be shown, some asteroids were identified as having Gaia reflectance spectra similar to the visible part of the spectrum of EC~002. The number of matches found with each method and each sample is recapitulated on Table \ref{tab:recap}, and the detail of the number and name of each asteroid matching and with which method it was found is given on Table~\ref{tab:bestmatches}.

\begin{table}[h!]
\begin{ThreePartTable}
\centering
\caption{Accepted asteroids as candidate matches for the different samples of EC~002, according to the method used. \label{tab:recap}}
\begin{tabular}{l c c c}
		\hline
		Sample & Spectral parameter & CM & Total \\ 
        \hline 
        Powder + slabs & 41  &  18 / 10  &  51 \\
        SW low  & 56  & 23 / 15 & 71 \\
        SW medium  & 12 &  8 / 5  & 17  \\
        SW high  & 2 &  1 / 1  & 3 \\
        \hline
\end{tabular}
\begin{tablenotes}
      \small
      \item[\bfseries Note:] CM stands for curve matching. The first number in the CM column is the number of asteroids found using the curve matching method for a given sample of the meteorite, and the second number corresponds to the number of asteroids not already found with the spectral parameter method. The total number of matches for each sample is indicated in the last column. 
    \end{tablenotes}
  \end{ThreePartTable}
\end{table}

\subsection{SDSS-Gaia spectral parameters comparison}

\begin{figure*}[ht]
  \centering
  \input{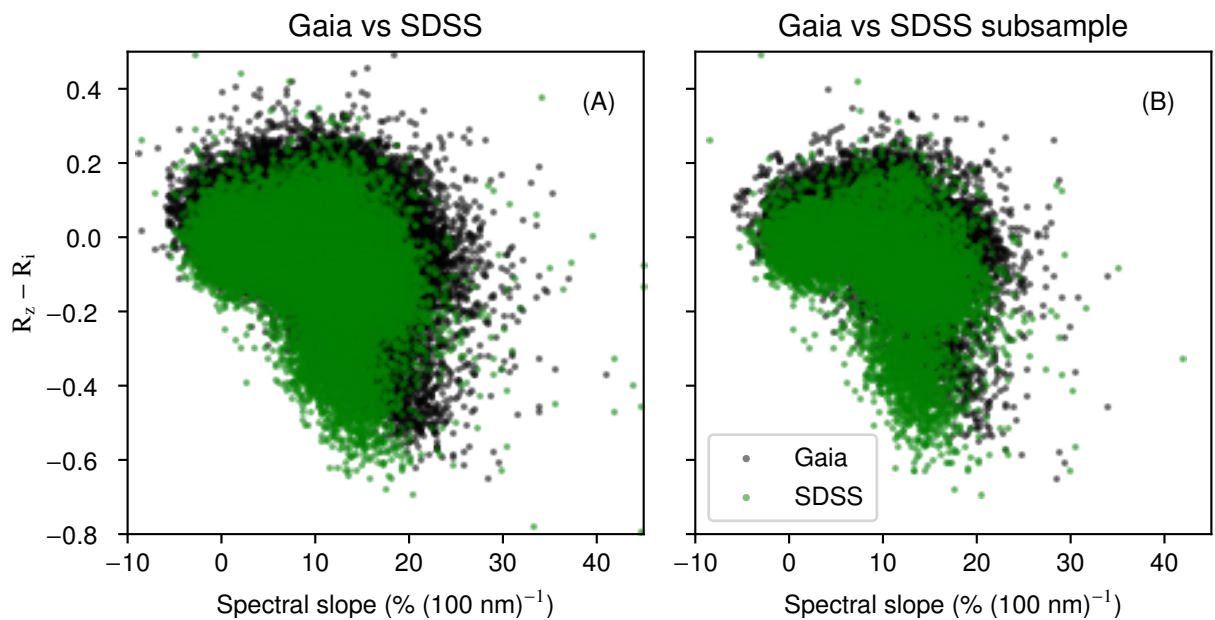}
    \caption{Comparison of the Band I depth $\mathrm{R_{z}-R_{i}}$ and the spectral slope for Gaia (black) and the SDSS (green) asteroids. Panel A: comparison between the 60\,518 asteroids of Gaia and the 33\,584 asteroids of the SDSS. Panel B: comparison between a sub-sample of 14\,129 asteroids both observed by the SDSS and Gaia. A shift in $\mathrm{R_{z}-R_{i}}$ between the spectra of Gaia and the SDSS is visible in both panels.}
    \label{fig:PlotgaiaSDSS}
\end{figure*}

\begin{figure*}[ht]
  \centering
  \input{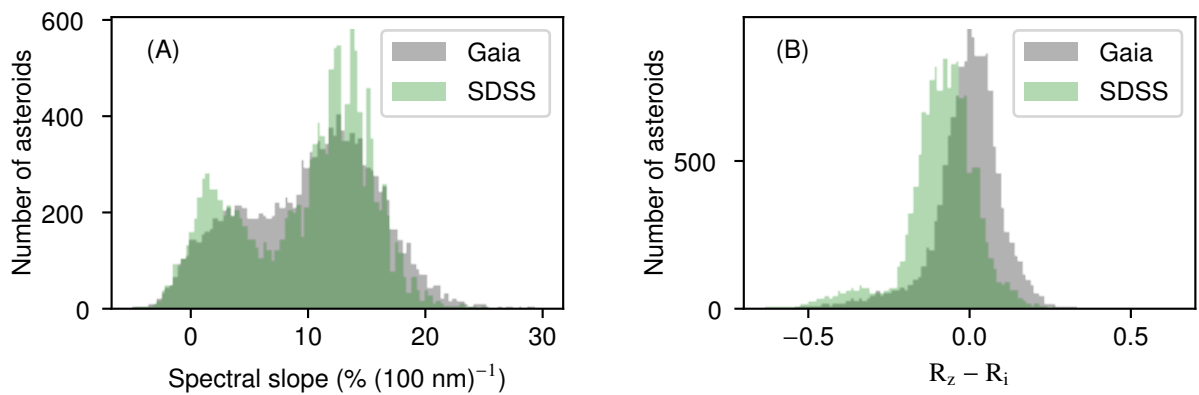}
    \caption{Comparison of the $\mathrm{R_{z}-R_{i}}$ parameter and of the spectral slope for the sub-sample of 14\,129 asteroids observed both by Gaia and SDSS. We note that the slope distribution (panel A) of Gaia has a wing that extends to redder slopes than the SDSS. Panel (B) shows that the distributions of $\mathrm{R_{z}-R_{i}}$ for Gaia and SDSS are clearly shifted, with Gaia seeing a less deep Band I compared to the SDSS. Gaia and SDSS $\mathrm{R_{z}-R_{i}}$ parameter histograms can be superimposed when a constant value of 0.07 is subtracted from Gaia $\mathrm{R_{z}-R_{i}}$-values. 
    \label{fig:histo-gaia-sdss}} 
\end{figure*}

First, because \cite{barrat2021} did not find satisfactory matches between EC~002 and the SDSS data, we started our analysis by investigating potential differences between Gaia and the SDSS dataset. To compare them, we calculated the spectral slope and the $\mathrm{R_{z}-R_{i}}$ parameter for every object in each dataset. The spectral slope was computed by linearly fitting the three SDSS reflectance data points in the g, r and i SDSS filters using a one-degree polynomial fit (numpy polynomial.polyfit). The same spectral parameters were calculated for the 60\,518 Gaia spectra as explained in Sect.~\ref{methods}.


As can be seen in Fig.~\ref{fig:PlotgaiaSDSS}, the two datasets appear to be shifted in the spectral parameter space. In order to study these apparent shifts, we used a sub-sample of 14\,129 asteroids having observations both in Gaia and SDSS. Figure \ref{fig:histo-gaia-sdss} shows histograms of the spectral parameters of this sub-sample, where a shift in $\mathrm{R_{z}-R_{i}}$ is clearly visible on panel (B). To evaluate the value of this shift, the median value of $\mathrm{R_{z}-R_{i}}$ was calculated for both datasets of the sub-sample and we found that Gaia data have 0.076 times higher $\mathrm{R_{z}-R_{i}}$ than the SDSS, doing the difference of the two. While there is a general agreement in spectral slope between the two surveys (difference between median slope of both surveys of only 0.52), a wing of higher slope values for Gaia asteroids can be noticed in Fig.~\ref{fig:histo-gaia-sdss}~(A), meaning that Gaia detects objects redder than the SDSS. The shift in $\mathrm{R_{z}-R_{i}}$ is quite significant and remains when considering only objects with a high S/N (S/N > 100 for example). We chose not to correct Gaia data from this shift in this work since we do not know its causes.


A potential reason for this shift could be the different choice of solar analogues between the two surveys. Indeed, a mean solar analogue spectrum was used to retrieve the asteroids spectra from Gaia, and solar colours are needed to convert colour indices to reflectances for the SDSS. It is possible that the accuracy of the solar analogues or the solar colours used is at the origin of this shift. This difference between the SDSS and Gaia will be investigated in future works.


\begin{figure*}[ht]
  \centering
    \begin{adjustbox}{clip,trim=0.8cm 0.5cm 1.8cm 1cm,max width=\textwidth}
  \input{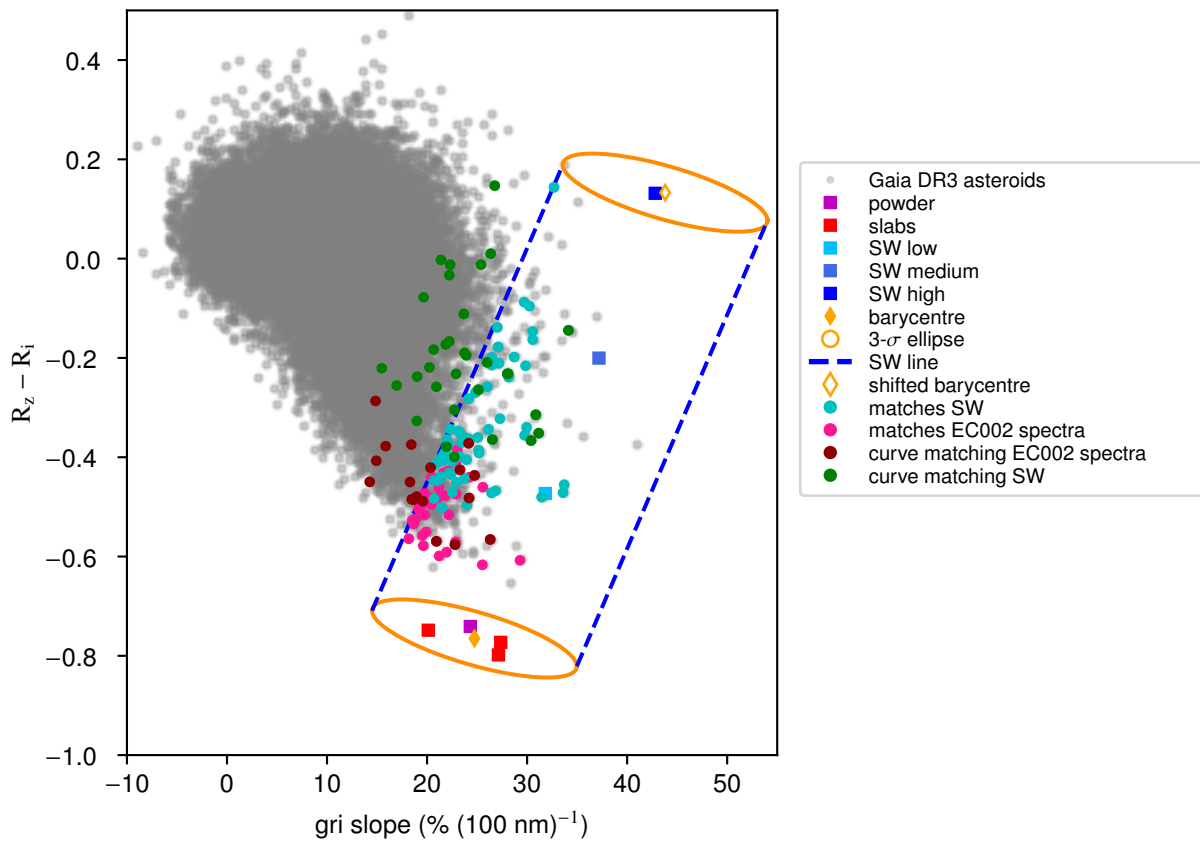}
  \end{adjustbox}
    \caption{ Distribution of depth of the 1~\SI{}{\micro\meter} band with respect to the spectral slope of every Gaia asteroid (grey dots). Red squares: raw slabs of the meteorite EC~002. Purple square : powder sample of the meteorite. Full orange diamond: barycentre of these 4 samples, and empty orange diamond: shifted barycentre along the `space weathering line'. The squares going from light blue to dark blue represent the modelled space-weathered spectra of EC~002, with different space weathering intensity. The orange ellipses are the 3-$\sigma$ ellipse respectively around the barycentre and shifted following the space weathering behaviour of EC~002. The two dashed blue lines delimit a `possible matches area', within which are represented asteroids matching EC~002. Pink dots: asteroids matching the raw slabs and powder sample of EC~002. Cyan dots: asteroids matching the space-weathered samples. Dark red dots: asteroids matching EC~002 by the curve matching method. Green dots: asteroids matching the space-weathered spectra of EC~002 by this same method.  \label{fig:zeta_xi_EC002}
    }
\end{figure*}

\subsection{Spectral parameter matching}

In the following are described the potential matches of EC~002 obtained with the study of spectral parameters. The slope and $\mathrm{R_{z}-R_{i}}$ spectral parameters were calculated for the spectra of the powder samples and all raw slab samples of EC~002 (see Table~\ref{tab:ec}). The average and standard deviation for the slope and Band I depth are of $ 24.7 \pm 2.9 $ \% (100~nm)$^{-1}$ and $ -0.76 \pm 0.02 $, respectively. In Fig.~\ref{fig:zeta_xi_EC002}, the corresponding point is plotted as an orange diamond. We can observe that the points corresponding to the different samples of the meteorite plots away from any group of asteroids in the spectral parameter space, as already observed in Fig.~S14 of the supplementary material of \cite{barrat2021}.

\begin{table}[h!]
\centering
\caption{Spectral slope and Band I depth evaluated for different samples of EC~002.\label{tab:ec}}
\begin{tabular}{l c c}
		\hline
		Sample & spectral slope (\%(100~nm)$^{-1}$) & $\mathrm{R_{z}-R_{i}}$ \\
		\hline
		 Powder & 24.3 & -0.74  \\
         Raw slab 1 & 27.4 & -0.77 \\
         Raw slab 2 & 20.1 & -0.75 \\
         Raw slab 3 & 27.1 & -0.80 \\
		\hline
\end{tabular}
\end{table}

In the spectral parameter space, using the average point as a centre, we defined a 3$\sigma$ confidence ellipse as described in section~\ref{methods} in which no asteroid is contained (Fig.~\ref{fig:zeta_xi_EC002}). As described in Sect.\ref{methods}, we calculated a `space weathering line' of equation: $y = 0.047 x - 1.92$. The coefficient of determination of this fit is $R^{2}$ = 0.986, meaning that the linear fit to the space weathering modelled spectra of \cite{barrat2021} is good. This line plots to the right side of the spectral parameter space, where only a few asteroids are present.

Using the parameters of the linear fit, we extended the 3$\sigma$ ellipse along the space weathering line, defining a `possible matches area' that contains 305 asteroids listed in Table~\ref{tab:SW}, which spectra were visually inspected.

Following several criteria we rejected $\sim$63.8\% of the sample. As a first step, we rejected the objects that have known VIS or NIR spectrum in the literature that allowed to distinguish them from EC~002. It corresponds to 10.2\% of the initial sample of 305 asteroids. Then, we removed the objects with more than three flagged bands in the Gaia spectrum (1.8\% of the sample). Finally, we rejected the asteroids that had either a too noisy spectrum, or that were visually different from the spectra of EC~002 in either BP or RP parts (51.8\% of the sample). For the last case we noticed that several spectra, otherwise similar to the one of EC~002 show a steep increase of the reflectance in the red part, making their Band I centre shifted compared to the one of the meteorite. We chose to reject such objects of the list of candidate matches.

After our visual inspection, 110 asteroids were retained (Table.~\ref{tab:SW}). Among these validated asteroids, 106 objects have been given a spectrum for the first time by the Gaia mission, and 41 asteroids were identified to have a reflectance spectrum similar to the laboratory spectra of EC~002. These objects are defined as matches. The matches of the four laboratory samples of the meteorite were not considered separately here, because these samples show almost indistinguishable spectra in the visible wavelength range. The spectra of the matches are shown on Fig.~\ref{fig:matchesnoSW}, and their median signal-to-noise ratio (S/N) is of 26.3.

In addition, there are 70 asteroids matching the space-weathered spectra of EC~002: 56 asteroids match the spectrum on which has been applied a low space weathering, 12 asteroids match the medium space-weathered spectrum and only two asteroids match the highly space-weathered spectra. Their spectra are shown respectively on Fig.~\ref{fig:matchesSWlow}, Fig.~\ref{fig:matchesSWmed} and Fig.~\ref{fig:matchesSWhigh}. The median S/N of the matches is of 23.0 for the low space-weathering of EC~002, of 18.2 for the median space weathering, and of 15.97 for asteroid (9974) Brody and 14.05 for asteroid (19754) Paclements.

\subsection{Curve matching}
We applied the curve matching method to the different laboratory spectra of the EC~002 (powder and slabs) and to the entire dataset of 60\,518 Gaia asteroid reflectance spectra. As mentioned before, the matches of the four laboratory samples were not considered separately here due to the similar visible spectra of these samples.

We considered only the cases giving $\chi_\text{red}^2$<2, resulting in a list of 58 bodies matching EC~002 listed on Table.~\ref{tab:slabPowderCM}. After a visual inspection of their spectra, several objects were rejected (Table~\ref{tab:slabPowderCM}). The final list of potential matches to EC~002 meteorite contains 18 asteroids. Amongst these, ten asteroids were not found with the spectral parameters method:
(16856) Banach,
(17056) Boschetti,
(54062) 2000GX135,
(63653) 2001QQ109,
(77147) 2001EV6,
(77935) 2002GM54,
(89556) 2001XS98,
(123113) 2000SH361,
(124884) 2001TE41, and
(164121) 2003YT1. The spectra of these ten bodies are shown on Fig.~\ref{fig:matchesnoSW_CM}.

The curve matching method was then applied to the space-weathered samples of EC~002. For the low space-weathered spectrum, 269 asteroids had a $\chi_\text{red}^2$<2. There was 223 asteroids with $\chi_\text{red}^2$<2 for the medium space-weathered spectrum, and only 12 asteroids for the highly space-weathered spectrum.

Most asteroids were rejected following the criteria exposed earlier. We finally found 23 asteroids as potential analogues of the low space-weathered EC~002, eight asteroids matching the medium space-weathered EC~002 and one asteroid matching the highly space-weathered meteorite. These objects are listed in Table~\ref{tab:SWCM}. The asteroids that were found as matches with this method and not with the spectral parameters method are asteroids (10131) Stanga, (15623) 2000 HU30, (18780) Kuncham, (20535) Marshburrows, (22276) Belkin, (22538) Lucasmoller, (32835) 1992EO5, (33423) 1999DK, (33852) Baschnagel, (33934) 2000LA30, (65504) 3544P-L, (74378) 1998XH11, (79827) 1998WU3, (100440) 1996PJ6, and (103308) 2000AH55 for the low SW ; asteroids (68089) 2000YS108, (68946) 2002PX138, (93797) 2000WO43, (108899) 2001PP5, (145532) 2006FD42 and (230762) 2003WP192 for the medium SW and asteroid (33809) 1999XK152 for the high SW. Their spectra are shown respectively on Fig.~\ref{fig:matchesSWlow_CM}, Fig.~\ref{fig:matchesSWmed_CM} and Fig.~\ref{fig:matchesSWhigh_CM}.


\onecolumn

\begin{table} 
\caption{Accepted asteroids as candidate matches for the different samples of EC~002. \label{tab:bestmatches}}
\begin{minipage}{0.5\textwidth}
\begin{tabular}{lc}
\toprule
Asteroid & Method \\
\midrule
        \textbf{Powder + raw slabs} \\ 
		\hline
        (1643) Brown & Spectral parameters \\
        (1946) Walraven & Spectral parameters \\
        (2432) Soomana & Spectral parameters \\
        (3188) Jekabsons & Spectral parameters \\
        (3651) Friedman & Spectral parameters \\
        (3869) Norton & Spectral parameters \\
        (4302) Markeev & Spectral parameters \\
        (5121) Numazawa & Spectral parameters \\
        (6853) Silvanomassaglia & Spectral parameters + CM \\
        (6876) Beppeforti & Spectral parameters \\
        (8587) Ruficollis & Spectral parameters \\
        (8827) Kollwitz & Spectral parameters \\
        (9197) Endo & Spectral parameters \\
        (9433) 1997 CF3 & Spectral parameters \\
        (10156) 1994 VQ7 & Spectral parameters + CM \\
        (10671) Mazurova & Spectral parameters \\
        (10902) 1997 WB22 & Spectral parameters \\
        (11155) Kinpu & Spectral parameters \\
        (12551) 1998 QQ39 & Spectral parameters \\
        (13839) 1999 XF29 & Spectral parameters \\
        (15989) 1998 XK39 & Spectral parameters \\
        (16856) Banach & CM \\ 
        (17056) Boschetti & CM \\ 
        (17240) Gletorrence & Spectral parameters \\ 
        (20454) Pedrajo & Spectral parameters + CM \\ 
        (24286) 1999 XU188 & Spectral parameters \\ 
        (24892) 1997 AD3 & Spectral parameters + CM \\ 
        (26573) 2000 EG87 & Spectral parameters \\ 
        (27262) 1999 XT184 & Spectral parameters \\ 
        (28162) 1998 VD14 & Spectral parameters \\ 
        (30769) 1984 ST2 & Spectral parameters \\ 
        (33418) Jacksonweaver & Spectral parameters \\ 
        (36431) 2000 PJ12 & Spectral parameters \\ 
        (44150) 1998 HC108 & Spectral parameters \\ 
        (47232) 1999 VQ36 & Spectral parameters \\ 
        (49101) 1998 RE76 & Spectral parameters \\ 
        (54062) 2000 GX135 & CM \\ 
        (55549) 2001 XC59 & Spectral parameters + CM \\ 
        (56904) 2000 QP171 & Spectral parameters \\ 
        (63653) 2001 QQ109 & CM \\ 
        (77147) 2001 EV6 & CM \\ 
        (77935) 2002 GM54 & CM \\ 
        (87093) 2000 LW6 & Spectral parameters \\ 
        (88955) 2001 TW42 & Spectral parameters + CM \\
        (89556) 2001 XS98 & CM \\ 
        (90604) 4813 P-L & Spectral parameters \\ 
        (123113) 2000 SH361 & CM \\ 
        (124884) 2001 TE41 & CM \\ 
        (164121) 2003 YT1 & CM \\ 
        (205560) 2001 SC282 & Spectral parameters + CM \\ 
        (310436) 2000 AB169 & Spectral parameters + CM \\
        \hline 
        \textbf{SW low} \\
        \hline 
        (4088) Baggesen & Spectral parameters \\ 
        (6003) 1988 VO1 & Spectral parameters \\ 
        (6789) Milkey & Spectral parameters \\ 
        (8243) Devonburr & Spectral parameters \\ 
        (8483) Kinwalaniihsia & Spectral parameters \\ 
\bottomrule
\end{tabular}

\end{minipage} \hfill
\begin{minipage}{0.5\textwidth}
\begin{tabular}{lc}
\toprule
Asteroid & Method \\ 
\midrule
        \textbf{SW low} \\
        \hline 
        (8692) 1992 WH & Spectral parameters \\ 
        (9753) 1990 QL3 & Spectral parameters \\ 
        (10131) Stanga & CM \\ 
        (11920) 1992 UY2 & Spectral parameters  \\ 
        (14511) Nickel & Spectral parameters \\ 
        (15088) Licitra & Spectral parameters \\ 
        (15623) 2000 HU30 & CM \\ 
        (17739) 1998 BY15 & Spectral parameters \\ 
        (17821) Bolsche & Spectral parameters \\ 
        (17882) Thielemann & Spectral parameters \\ 
        (17943) 1999 JZ6 & Spectral parameters \\ 
        (18344) 1989 TN11 & Spectral parameters \\ 
        (18780) Kuncham & CM \\ 
        (19978) 1989 TN6 & Spectral parameters \\ 
        (20289) Nettimi & Spectral parameters \\ 
        (20535) Marshburrows & CM \\ 
        (21318) 1996 XU26 & Spectral parameters \\ 
        (22276) Belkin & CM \\ 
        (22538) Lucasmoller & CM \\ 
        (23766) 1998 MZ23 & Spectral parameters \\ 
        (24569) 9609 P-L & Spectral parameters \\ 
        (24684) 1990 EU4 & Spectral parameters + CM \\ 
        (26084) 1981 EK17 & Spectral parameters \\ 
        (26851) Sarapul & Spectral parameters \\ 
        (27876) 1996 BM4 & Spectral parameters + CM \\ 
        (27884) 1996 EZ1 & Spectral parameters \\ 
        (28132) Karenzobel & Spectral parameters \\ 
        (29171) 1990 QK3 & Spectral parameters \\ 
        (30426) Philtalbot & Spectral parameters \\ 
        (30834) 1990 VR6 & Spectral parameters \\ 
        (32835) 1992 EO5 & CM \\ 
        (33423) 1999 DK & CM \\ 
        (33852) Baschnagel & CM \\ 
        (33934) 2000 LA30 & CM \\ 
        (33947) 2000 ML1 & Spectral parameters + CM \\ 
        (35364) Donaldpray & Spectral parameters \\ 
        (39940) 1998 FR99 & Spectral parameters \\ 
        (41894) 2000 WH121 & Spectral parameters \\ 
        (43278) 2000 ES109 & Spectral parameters + CM \\ 
        (44162) 1998 HC148 & Spectral parameters \\ 
        (45787) 2000 OJ24 & Spectral parameters \\ 
        (48039) 2001 DT69 & Spectral parameters \\ 
        (51659) Robohachi & Spectral parameters \\ 
        (53417) 1999 NP38 & Spectral parameters \\ 
        (53661) 2000 DU62 & Spectral parameters \\ 
        (53899) 2000 FM49 & Spectral parameters \\ 
        (56561) Jaimenomen & Spectral parameters + CM \\ 
        (58640) 1997 WH18 & Spectral parameters \\ 
        (61098) 2000 LY28 & Spectral parameters \\ 
        (64458) 2001 VF35 & Spectral parameters \\ 
        (65504) 3544 P-L & CM \\ 
        (74107) 1998 QM37 & Spectral parameters \\ 
        (74378) 1998 XH11 & CM \\ 
        (75323) 1999 XY47 & Spectral parameters \\ 
        (79827) 1998 WU3 & CM \\ 
        (87216) 2000 OG38 & Spectral parameters \\ 
        (89776) 2002 AL90 & Spectral parameters \\ 
\bottomrule
\end{tabular}
\end{minipage}
\end{table}

\twocolumn

\begin{table}
\renewcommand\thetable{3}
\caption{continued.}
\begin{ThreePartTable}
\centering
\begin{tabular}{l c}
		\hline
		Asteroid & Method \\ 
		\hline
        \textbf{SW low} \\ 
        \hline 
		(89952) 2002 JB20 & Spectral parameters + CM \\ 
        (92593) 2000 PN16 & Spectral parameters \\ 
        (100440) 1996 PJ6 & CM \\ 
        (103308) 2000 AH55 & CM \\ 
        (108139) 2001 GL11 & Spectral parameters + CM\\ 
        (112326) 2002 MM4 & Spectral parameters + CM \\ 
        (122122) 2000 JM16 & Spectral parameters \\ 
        (128450) 2004 NX24 & Spectral parameters \\ 
        (134916) 2000 YP53 & Spectral parameters \\ 
        \hline 
        \textbf{SW medium} \\ 
        \hline
        (13133) Jandecleir & Spectral parameters \\ 
        (18143) 2000 OK48 & Spectral parameters \\ 
        (31060) 1996 TB6 & Spectral parameters \\ 
        (42822) 1999 NT13 & Spectral parameters + CM \\ 
        (44322) 1998 RZ42 & Spectral parameters + CM \\ 
        (49141) 1998 SM41 & Spectral parameters \\ 
        (51379) 2001 BY7 & Spectral parameters \\ 
        (52408) 1993 TJ34 & Spectral parameters \\ 
        (68089) 2000 YS108 & CM \\ 
        (68946) 2002 PX138 & CM \\ 
        (90843) 1995 YZ22 & Spectral parameters \\ 
        (93797) 2000 WO43 & CM \\ 
        (99714) 2002 JQ41 & Spectral parameters \\ 
        (108899) 2001 PP5 & CM \\ 
        (122125) 2000 JO17 & Spectral parameters \\ 
        (145532) 2006 FD42 & CM \\ 
        (230762) 2003 WP192 & Spectral parameters + CM \\
        \hline 
        \textbf{SW high} \\ 
        \hline
        (9974) Brody & Spectral parameters \\ 
        (19754) Paclements & Spectral parameters \\ 
        (33809) 1999 XK152 & CM \\
        \hline
\end{tabular}
\begin{tablenotes}
      \small
      \item[\bfseries Note:] Asteroids showing a spectrum matching the powder and raw slabs samples of the meteorite are 51 in number, while 71 asteroids match the low space-weathered spectrum of EC~002, 17 asteroids match its medium space-weathered spectrum and three asteroids match its highly space-weathered spectrum. CM stands for curve matching. 
    \end{tablenotes}
  \end{ThreePartTable}
\end{table}


\begin{figure*}[!h]
  \centering
      \begin{adjustbox}{clip,trim=1cm 1cm 1.8cm 2cm,max width=\textwidth}
  \input{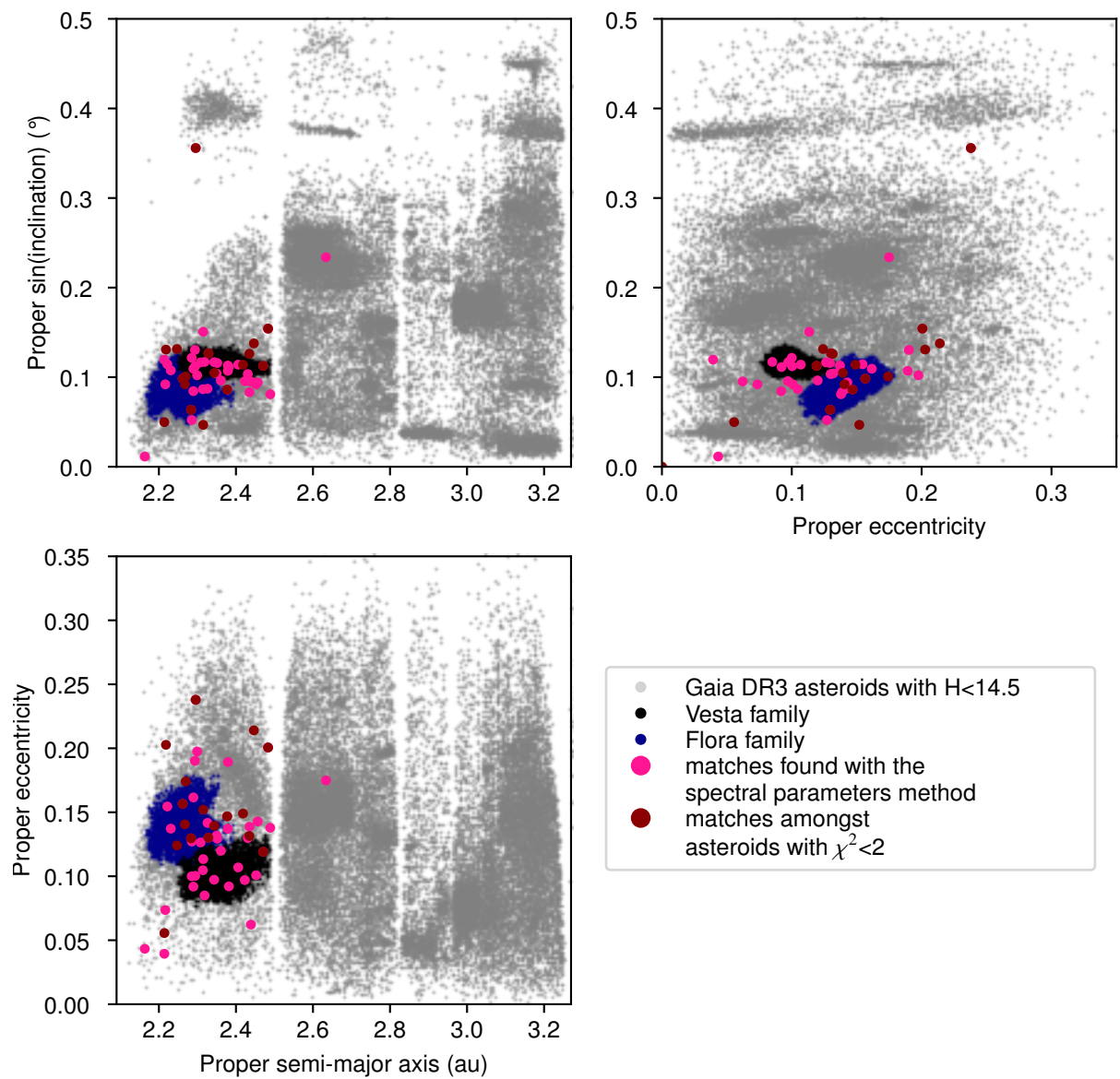}
  \end{adjustbox}
  \caption[Proper orbital element ]{Proper orbital element\footnotemark  plots of Gaia asteroids with absolute magnitude H$<$14.5 (light grey dots). Vesta and Flora family members are indicated respectively with black and blue dots. Asteroids that were found to be spectroscopically matching with EC~002 after visual inspection are plotted with dots circles indicated in the legend above. \label{fig:a-sini-e-core}}
\end{figure*}

\section{Discussion}
\label{discussion}

\begin{figure*}[h!]
  \centering
        \begin{adjustbox}{clip,trim=1cm 1cm 0.5cm 2cm,max width=\textwidth}
  \input{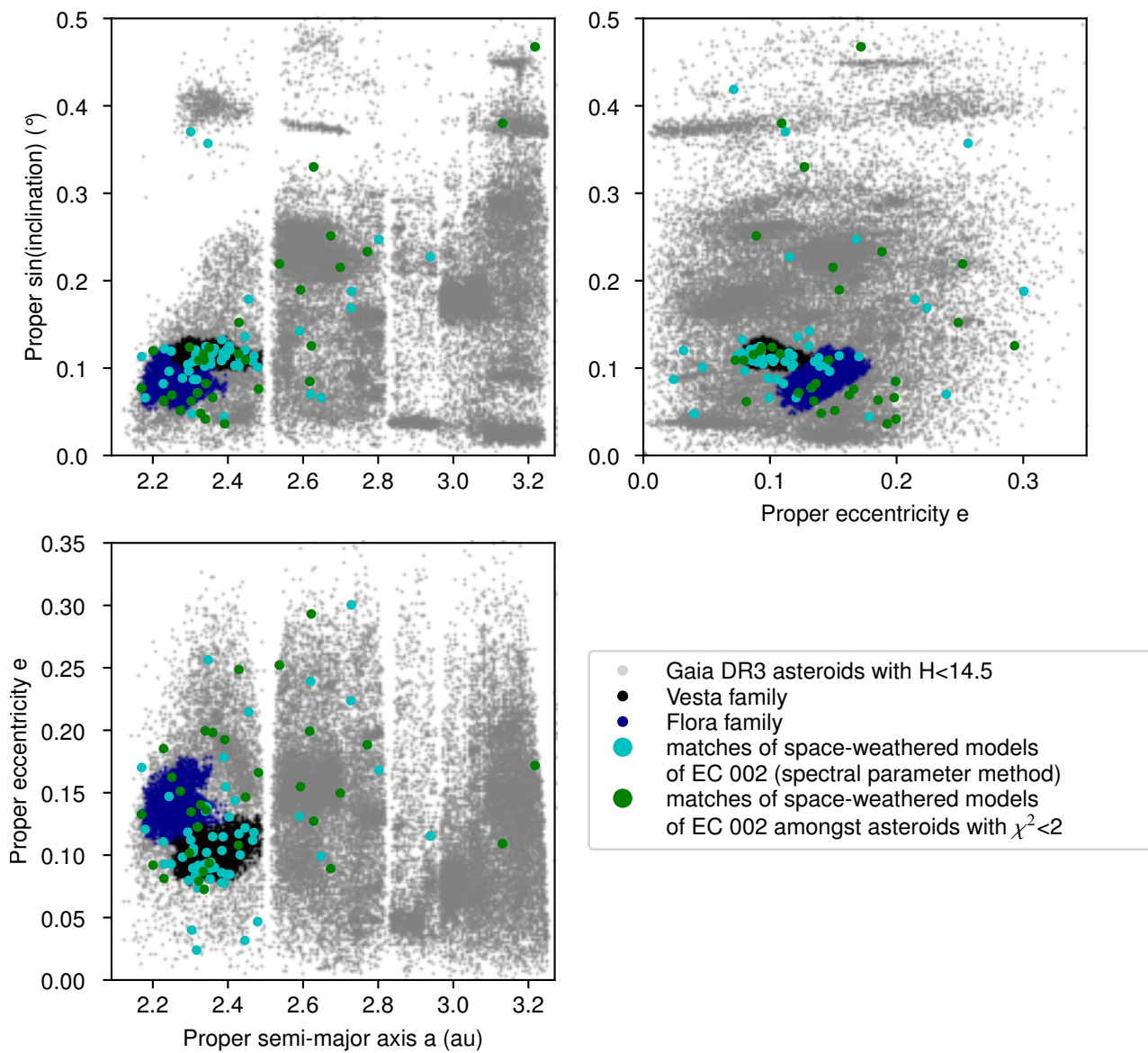}
  \end{adjustbox}
\caption{\label{fig:a-sini-e-sw} Same as Fig.~\ref{fig:a-sini-e-core}, but with cyan dots indicating asteroids that have spectral parameters compatible with the space-weathered models of EC~002 derived by \cite{barrat2021} in the spectral parameters space. The green dots are asteroids matching the weathered models of the meteorite spectra using the curve matching method. } 
\end{figure*}

Asteroids spectroscopically matching EC~002 are extremely rare. We find only 51 asteroids matching the non-space-weathered spectrum of EC~002, and 91 asteroids matching its spectrum on which was modelled the effect of space weathering to various degrees. Considering the entire Gaia sample of over 60\,518 Solar System minor bodies, it means a mere 0.08\% of the sample for the non-space-weathered samples and 0.15\% of the sample for the space-weathered EC~002. This confirms the conclusions of \cite{barrat2021} about the scarcity of analogues of EC~002 among the asteroid population.

The best matches of the different samples of EC~002 are defined as the objects found using both methods. For the four laboratory samples, there are seven best matches:
(6853) Silvanomassaglia,
(10156) 1994VQ7,
(20454) Pedrajo,
(55549) 2001XC59,
(88955) 2001TW42,
(205560) 2001SC282, and (310436) 2000AB169. For the spectra on which was applied a low space weathering model, there are eight best matches: asteroids (24684) 1990 EU4,
(27876) 1996BM4,
(33947) 2000ML1,
(43278) 2000ES109,
(56561) Jaimenomen,
(89952) 2002JB20,
(108139) 2001GL11, and (112326) 2002MM4. For the medium space weathering model, asteroids
(42822) 1999NT13,
(44322) 1998RZ42,
and (230762) 2003WP192 are found by both methods. No asteroid is found by both methods for the highly space-weathered spectrum.

Both methods give quite different asteroid as matches. Indeed, we did not filter the objects considering their S/N but we noticed that the large majority of objects retained as potential analogues of EC~002 have S/N-values lower than a hundred. For the curve matching method, the median value of the S/N for the accepted objects is of 17.2. This low value is explained by the fact that the chosen curve matching parameter favours observations with large error bars, hence low S/N observations. This method thus filters out objects with higher S/N found by the spectral parameter method that appear to be very good matches by visual inspection, such as asteroid (5121) Numazawa. In order to evaluate the results of this curve matching method, we performed some tests with an alternative Least Squares method. We computed the sum of the squared residuals between the meteorite and the asteroids spectra, removing the flagged points as was done for the $\chi^2_{red}$ calculation and not taking the uncertainties into account. The sum of the squared residuals used is described in Eq.~\ref{eq:LS}:
\begin{equation}
    \mathrm{R^2 =  \sum_{i}^{N} {(A_{i}-f . M_{i})^{2}}},
    \label{eq:LS}
\end{equation}
with $M_{i}$ the meteorite spectrum, $A_{i}$ a Gaia asteroid reflectance spectrum and $f$ a normalisation factor similar to the one in Eq.~\ref{formula:f} but without consideration of the uncertainties. This method spans a wider range of S/N, giving only potential matches of EC~002 among asteroids with S/N above 25 for the powder sample of the meteorite for example. Most asteroids found as potential matches plot inside the `possible matches area' of Fig.\ref{fig:zeta_xi_EC002} and thus were already found with the spectral parameters method. This method is therefore a complement of the spectral parameter method, but since it is sensitive to outliers it requires a visual inspection as well to assess the quality of the matches. The curve matching method with the $\chi_{red}^2$ has the advantage that it explores a larger area in the spectral parameter space, finding objects outside the `possible matches area' even though they are low S/N asteroids. These objects should be further observed and studied in future analysis to evaluate how good a match they are.

All matched asteroids with the non-space-weathered meteorite spectra are located in the inner part of the main belt (Fig.~\ref{fig:a-sini-e-core}), between the secular resonance $\nu_6$ and the 3:1 mean motion resonance with Jupiter. The asteroids matched with the space-weathered meteorite are more scattered across the main belt, even though most objects can be found in the inner main belt as well, in particular close to the Vesta family (Fig.~\ref{fig:a-sini-e-sw}). 

Some of the asteroids matching the different samples of EC~002 are members of known collisional families, according to the membership of \cite{nesvorny2015}. Among the 142 matches of the different samples of EC~002, 23.9\% of the asteroids belong to the Vesta family, 9.8\% belong to the Flora family and a mere 7.7\% belong to other families (mainly Nysa-Polana).

The presence of asteroids matching with EC~002 inside the Vesta family could be due to two possible reasons. If these objects are real analogues to EC~002, they could be interlopers inside the Vesta family since EC~002 is chemically distinct from the HEDs \citep{barrat2021}.
The second possibility is that these asteroids are true Vesta family members compositionally alike HEDs, but which happen to have a Gaia reflectance spectrum in the visible range very similar to that of EC~002. In fact, the distinction between the reflectance spectra of EC~002 and HEDs in the visible is difficult, and relies mainly on the position of the Band I centre. However, the last Gaia bands may show a fast increase in reflectance due to light contamination in the RP \citep{galluccio2022}. This contamination could result in a shift of the Band I centre. Future analysis of the Gaia reflectance spectra could help solve this issue, and future near-infrared spectroscopy of these bodies might be able to reveal if they are more similar to EC~002 or to the HEDs and to Vesta family members.

\footnotetext{Data retrieved from the Belgrade catalogue \url{http://asteroids.matf.bg.ac.rs/fam/properelements.php}.}

The Flora family is a large collisional family adjacent to the $\nu_6$ \citep{nesvorny2015}, which is the most efficient region to deliver main-belt asteroids to Earth-crossing orbits \citep{Granvik2016}. Hence, this family could be an important source of near-Earth asteroids \citep[]{LaSpina2004, Kryszczynska2013} and meteorites \citep{Nesvorny2002}. The Flora family is mainly constituted of S-type asteroids \citep[][and references therein]{Oszkiewicz2015, nesvorny2015}, which are linked to ordinary chondrites. \cite{barrat2021} showed that EC~002 could be derived from the partial melt of a planetesimal of non-carbonaceous chondritic composition, which experienced heating during its accretion and consequently formed an igneous crust. If the asteroids belonging to the Flora family are (i) real EC~002 analogues, and (ii) true members of the Flora family, this would confirm the spectroscopic diversity within this family pointed out by several studies \citep[][and references there in]{Oszkiewicz2015}. In addition, this would potentially point towards a differentiation of the family parent body, as has been already proposed \citep{Oszkiewicz2015}.

Given the very low number of asteroids belonging to the other asteroid families, it is difficult to assume that the parent body of EC~002 was part of these families. Every other asteroids potentially matching with EC~002 do not belong to known families. However, current family catalogues are based on algorithms that determine the membership of an object to a family based on its proper orbital elements only, in order to distinguish the different families and to clearly identify their cores \citep{nesvorny2015, tsirvoulis2018}. As a result, some background objects are designated as family members but are in fact interlopers, and some real family members are not considered as part of the family, leading to halos of asteroids surrounding known families \citep[]{parker2008, Broz2013}. In addition, the study of the dynamical behaviour of family and non-family asteroids by \cite{Dermott2018} lead to the conclusion that most asteroids in the inner main belt are or were originally part of the main known families, showing evidence that the families are very dispersed. Some of these dispersed families have been detected \citep[]{delbo2017,delbo2019} using the so-called V-shape method \citep{bolin2017}, and more families are probably left to be identified \citep{delbo2017,delbo2019,Dermott2018}.
Hence, it is possible that the asteroids matching with EC~002 that are not listed as family members are part of very old families of the inner main belt that escaped identification up to now. 

It is also established that laboratory spectra of meteorites do not necessarily match with the spectra of asteroids of analogue composition \citep[][and references therein]{brunetto2015}. The reason is that the reflectance spectra of asteroids are affected by the exposure of their surface to weathering agents in space, such as solar wind ions and micrometeorites. Space weathering models have been developed, for example by \cite{Hapke2001} or \cite{Brunetto2006}, in order to correct reflectance spectra from space weathering. The Hapke model is based on the calculation of the absorption coefficient of a silicate host medium in which small nanophase iron spheres are included \citep[see][for further details]{Brunetto2007}. The inclusion of nanophase iron inside a siliceous material changes its physical properties and alters its visible and infrared spectrum: the spectral slope is reddened, the silicate bands become shallower and less recognisable and the albedo of the object is darkened. However, the silicate band centres are not (or very little) affected by this type of space weathering \cite{gaffey2002}. This model was developed by studying the space weathering of the Moon and it successfully recreates it. For its appliance to an object to be relevant, the mineralogy of the object needs to be dominated by silicates with grains larger than the wavelength (Pierre Beck, private communication). Thus, the Hapke and other similar models \citep{Brunetto2006} have been applied to ordinary chondrites and allowed to link this type of meteorites to S-type asteroids, giving precious information about the mineralogy of these asteroids.
EC~002 is an achondrite with andesitic composition, corresponding to the partial melt of an ordinary chondrite and which contains silicates with large grains \citep{barrat2021}. Therefore, using the Hapke model makes sense to simulate the effect of space weathering on an asteroid of the same composition as EC~002, as what was implemented by \cite{barrat2021}.

\begin{figure}[h!]
  \centering
  \input{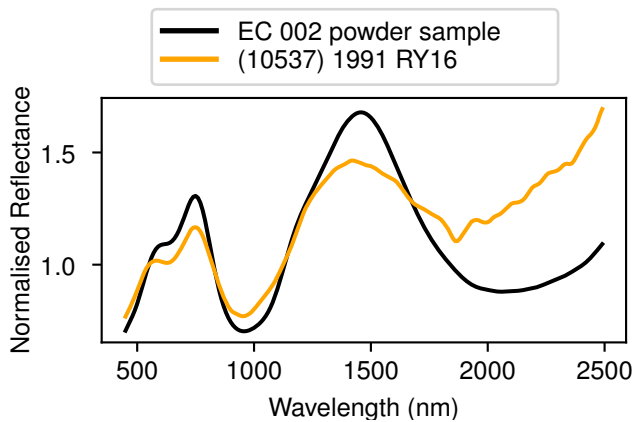}
    \caption{\label{fig:10537_EC002} VISNIR spectra of the powder sample of EC~002 (black lines) and of asteroid (10537) 1991~RY16 (orange lines) retrieved from Fig.1 of \cite{moskovitz2008}. Both spectra were normalised at 550~nm. The two spectra show a similar shape, they both show a band around 0.65~\SI{}{\micro\meter} and similar Band I centre. However their Band II centre is shifted with respect to each other.} 
\end{figure}

A characteristic feature of the EC~002 reflectance spectrum is the presence of an absorption band at 0.65~\SI{}{\micro\meter}. Unfortunately, this feature cannot be used as an absolute diagnostic feature in Gaia asteroid spectra for two reasons. First, the BP and RP spectra overlap in the region of this band. Since the spectrophotometers are independently calibrated \citep{DeAngeli2022}, their overlapping region can be affected by artefacts \citep{galluccio2022} and must be handled with care. The second reason is that some V-type asteroids also display an absorption band near 0.65~\SI{}{\micro\meter}, such as the asteroid (10537) 1991~RY16 \citep{moskovitz2008}. Interestingly, the visible reflectance spectra of asteroid (10537) 1991~RY16 and EC~002 are quite similar. This asteroid has already been found the closest match to the ungrouped achondrite (NWA)~7325 by \cite{cloutis2018} based on the spectral features of both bodies, without being a satisfactory match. In the near-infrared however, EC~002 shows a deeper Band II depth and a Band II centre at longer wavelengths (1.89~\SI{}{\micro\meter} for asteroid (10537) 1991~RY16, vs. 2.08~\SI{}{\micro\meter} for EC~002, as visible Fig.~\ref{fig:10537_EC002}). This points to the necessity of using near-infrared spectroscopy to distinguish potential EC~002 visible matches against V-type asteroids. 

\section{Conclusion and perspectives}

We searched for analogues of the andesitic meteorite EC~002 among the asteroid population, using Gaia visible reflectance spectra. We studied four different laboratory samples of the meteorite: three raw slabs and one powder spectrum; and three modelled space-weathered spectra of EC~002 were also analysed. As first method, we evaluated and compared the spectral parameters of each sample of the meteorite with the ones of the asteroids, studying the slope and the Band I depth of each asteroid spectrum. The second method used was a curve matching method.

For both methods, a visual inspection of the asteroid spectra and a search in the literature for already existing VIS and NIR spectra of these objects allowed us to deduce which asteroids are the most probable analogues of EC~002 among the main-belt asteroid population. The spectral parameter method gave 41 objects as potential analogues to the laboratory samples of EC~002, and 70 objects matching the space-weathered spectra of EC~002. These objects are mostly located in the inner main belt, around the Vesta and Flora families.

The curve matching method gave 18 objects matching the laboratory samples of the meteorite, also concentrated in the inner main belt. The curve matching with the modelled space-weathered spectra of the meteorite gave 23 asteroids as potential analogues of the low space-weathered EC~002, eight asteroids matching the medium space-weathered meteorite and only one asteroid matching the highly space-weathered EC~002. Because of the $\chi_{red}^2$ parameter used, only objects with a low S/N were found with this method. In the end, a total of 51 asteroids were found as potential analogues of the not-space-weathered EC~002, and 91 asteroids were found matching the modelled space-weathered spectra of the meteorite.

Finally, only 0.08\% of Gaia asteroids were found to be matching the laboratory samples of the meteorite, and 0.15\% were found matching the modelled space-weathered spectra. However, acquiring and studying the near-infrared spectra of these objects could help determining if they are real analogues of EC~002. If they are, they would be remnants of the original population of planetesimals that appeared in the early times of the Solar System and that showed an andesitic - and not basaltic - crust after differentiation. Traces of this original population would thus still exist in the main belt. Moreover, a full VISNIR spectrum would allow the study of more spectral parameters (Band II centre and band area ratio), which would give great clues about the quality of the matches presented in this work.

\begin{acknowledgements}
MG and MD acknowledge financial support from CNES and the Action Specifique Gaia.
MD, CA, and LG acknowledge financial support from the ANR ORIGINS (ANR-18-CE31-0014).
This work has made use of data from the European Space Agency (ESA) mission
{\it Gaia} (\url{https://www.cosmos.esa.int/gaia}), processed by the {\it Gaia}
Data Processing and Analysis Consortium (DPAC,
\url{https://www.cosmos.esa.int/web/gaia/dpac/consortium}). Funding for the DPAC
has been provided by national institutions, in particular the institutions
participating in the {\it Gaia} Multilateral Agreement. 
This work is based on data provided by the Minor Planet Physical Properties Catalogue (MP3C) of the Observatoire de la Côte d'Azur.
Spectra from \cite{barrat2021} were used as well in this study. The authors are grateful to Jean-Alix Barrat and Pierre Beck for their precious help.
\end{acknowledgements}


\bibliographystyle{aa} 
\bibliography{newbib.bib} 


\begin{appendix}

\clearpage
\onecolumn
\section{Tables of asteroids matching the spectra of EC~002}

{\begin{ThreePartTable}
\setlength\tabcolsep{3.5pt} 
\footnotesize\setlength{\tabcolsep}{2.6pt}
\begin{TableNotes}
  \item[\bfseries Note: ] The information in the table are the number and name of the 305 asteroids, if they are accepted or not as a match for EC~002 (1 if accepted, 0 if not), some notes about the visual inspection, the spectral type of the asteroid if determined and the method and relevant references associated (Ref column). The taxonomic scheme used for the type of each asteroid is the one used in the reference papers associated. Spec. stands for Spectroscopy and Phot. for Photometry.
\end{TableNotes}
\begin{longtable}{l c c c c c}
\caption{Asteroids within the 'possible matches area'.  \label{tab:SW}} \\
\hline 
Asteroid & Acceptance & Notes & Type & Method & Ref. \\ 
\hline 
\endfirsthead 
{\tablename\ \thetable\ -- continued.} \\ 
\hline 
Asteroid & Acceptance & Notes & Type & Method & Ref \\ 
\hline 
\endhead 
\hline 
\endfoot
\hline
\insertTableNotes
\endlastfoot
(289) Nenetta   & 0 &  longer band I centre   & A &   Spec. VISNIR   &  5  \\
(863) Benkoela  & 0 &  longer band I centre  & A &  Spec. VISNIR  &  5 \\
(956) Elisa  & 0 &  -   &  V  &  Spec. VIS and NIR  &  3, 6 \\
(1459) Magnya  & 0 &  VISNIR different  &  V  &  Spec. VISNIR  &  5 \\
(1468) Zomba & 0 & NIR different & V & Spec. VISNIR & 20 \\
(1488) Aura & 0 & different red part & A & Phot. VIS & 24 \\
(1643) Brown & 1 & - & S & Phot. VIS & 24 \\
(1709) Ukraina & 0 & A type spectrum & A & Spec. VISNIR & 25 \\
(1908) Pobeda  & 0 &  longer band I centre  &  S  &  Phot. VIS  &  24 \\
(1946) Walraven  & 1 &  -   &  V  &  Spec. VIS  &  4 \\
(2168) Swope  & 0 &  -   &  V  &  Spec. NIR  &  16,22 \\
(2371) Dimitrov & 0 & NIR different & V & Spec. VIS and NIR & 2, 6 \\
(2432) Soomana  & 1 &  -   &  V  &  Phot. VIS  &  10, 23 \\
(2442) Corbett & 0 & shorter band I centre & V & Spec. VISNIR & 5  \\
(2557) Putnam  & 0 &   shorter band I centre  &  S  &  Phot. VIS  &  24 \\
(2851) Harbin  & 0 &  -   &  V  &  Spec. VISNIR  &  5 \\
(2912) Lapalma  & 0 &  -   &  V  &  Spec. VISNIR  &  5 \\
(3104) Durer & 0 & different red part & K & Spec. VISNIR & 25\\
(3155) Lee  & 0 &  shorter band I centre  &  V  &  Spec. VISNIR  &  5 \\
(3188) Jekabsons  & 1 &  -   &  V  &  Spec. VIS  &  21 \\
(3651) Friedman  & 1 &  bad two last points   &  V  &  Phot. VIS  &  24 \\
(3817) Lencarter  & 0 &  shorter band I centre  &  -  &  -  &  - \\
(3869) Norton  & 1 &  article: related to 4 Vesta   &  V   &  Spec. VIS  &  1 \\
(3882) Johncox  & 0 &  -   &  V  &  Spec. VISNIR  &  18 \\
(4055) Magellan  & 0 &  -   &  V  &  Spec. VISNIR  &  5 \\
(4088) Baggesen  & 1 &  no clear 0.65~\SI{}{\micro\meter} band - SW low   &  -  &  -   &  - \\
(4302) Markeev  & 1 &  -   &  V  &  Phot. VIS  &  23 \\
(4402) Tsunemori  & 0 &   different band I shape  &  A  &  Spec. VISNIR  &  25 \\
(4692) SIMBAD  & 0 &  shorter band I centre  &  V  &  Phot. VIS  &  10, 23 \\
(5037) Habing & 0 & shorter band I centre & V & Spec. VISNIR & 25 \\
(5121) Numazawa  & 1 &  -  &  S  &  Phot. VIS  &  24 \\
(5498) Gustafsson  & 0 &  linked to howardites   &  V  &  Spec. VIS and NIR  &  6, 8 \\
(5696) Ibsen & 0 & different red part &  &   &  \\
(6003) 1988 VO1  & 1 &  SW low  &  X  &  Phot. VIS  &  24 \\
(6046) 1991 RF14  & 0 &   shorter band I centre  &  V  &  Spec. VISNIR  &  18 \\
(6159) Andreseloy  & 0 &  shorter band I centre  &  V  &  Spec. VISNIR  &  25 \\
(6369) 1983 UC  & 0 &  shorter band I centre  &  -  &  -   &  - \\
(6584) Ludekpesek  & 0 &  shorter band I centre  &  V  &  Phot. NIR  &  17 \\
(6728) 1991 UM  & 0 &  shorter band I centre  &  -  &  -   &  - \\
(6789) Milkey  & 1 &  SW low  &  -  &  -  &  - \\
(6853) Silvanomassaglia  & 1 &  -  &  V  &  Phot. NIR  &  17 \\
(6876) Beppeforti  & 1 &  -  &  S  &  Phot. VIS  &  24, 24 \\
(6877) Giada & 0 & shorter band I centre & V & Phot. NIR & 17 \\
(6964) Kunihiko  & 0 &  shorter band I centre  &  V  &  Phot. VIS  &  24 \\
(7294) Barbaraakey & 0 & flatter red part & S & Phot. VIS & 24 \\
(7529) Vagnozzi & 0 & flatter red part & V & Phot. VIS & 24 \\
(7889) 1994 LX & 0 & noisy & V & Spec. VISNIR & 20 \\
(7933) Magritte  & 0 &  shorter band I centre  &  X  &  Phot. VIS  &  24 \\
(7942) 1991 OK1  & 0 &  shorter band I centre  &  S  &  Phot. VIS  &  24 \\
(8031) Williamdana  & 0 &  shorter band I centre  &  V  &  Phot. VIS  &  24 \\
(8243) Devonburr  & 1 &  SW low  &  S  &  Phot. VIS  &  24, 24 \\
(8483) Kinwalaniihsia & 1 & SW low (without first and last bands) & V & Phot. VIS & 24 \\
(8587) Ruficollis  & 1 &  -  &  K  &  Phot. VIS  &  24 \\
(8660) Sano  & 0 &  longer band I centre  &  S  &  Phot. VIS  &  24 \\
(8669) 1991 NS1  & 0 &  shorter band I centre  &  S  &  Phot. VIS  &  24 \\
(8692) 1992 WH & 1 & SW low & S & Phot. VIS & 24 \\
(8827) Kollwitz & 1 & - & C & Phot. VIS & 24 \\
(8838) 1989 UW2  & 0 &   longer band I centre  &  A  &  Spec. VISNIR  &  19 \\
(9115) Battisti  & 0 &  shorter band I centre  &  V  &  Phot. VIS  &  24 \\
(9197) Endo  & 1 &  not very good VISNIR literature spectrum   &  V  &  Spec. VISNIR  &  22 \\
(9432) Iba & 0 & shorter band I centre & V & Phot. VIS & 24 \\
(9433) 1997 CF3  & 1 &  -  &  C  &  Phot. VIS  &  24 \\
(9593) 1991 PZ17  & 0 &  bump instead of 0.65~\SI{}{\micro\meter} band   &  S  &  Phot. VIS  &  24 \\
(9752) 1990 QZ1  & 0 &  longer band I centre  &  S  &  Phot. VIS  &  24 \\
(9753) 1990 QL3  & 1 &  SW low  &  -  &  -  &  - \\
(9974) Brody  & 1 &  SW high  &  -  &  -  &  - \\
(10156) 1994 VQ7  & 1 &  bad three last points  &  V  &  Phot. VIS  &  24 \\
(10319) Toshiharu  & 0 &   shorter band I centre V  &  V  &  Phot. VIS  &  23, 24 \\
(10418) 1998 WZ23  & 0 &  shorter band I centre  &  V  &  Phot. VIS  &  24 \\
(10438) Ludolph  & 0 &  shorter band I centre  &  -  &  -  &  - \\
(10578) 1995 LH  & 0 &  bad BP RP overlapping  &  -  &  -  &  - \\
(10671) Mazurova  & 1 &  -  &  S  &  Phot. VIS  &  24 \\
(10811) Lau & 0 & flatter red part & - & - & - \\
(10902) 1997 WB22 & 1 & - & - & -  & - \\
(11041) Fechner  & 0 &  shorter band I centre  &  V  &  Phot. VIS  &  7 \\
(11155) Kinpu  & 1 &  -  &  S  &  Phot. VIS  &  24 \\
(11764) Benbaillaud  & 0 &  shorter band I centre  &  V  &  Spec. VIS  &  8 \\
(11861) Teruhime  & 0 &   longer band I centre  &  -  &  -   &  - \\
(11890) 1991 FF  & 0 &   longer band I centre   &  C  &  Phot. VIS  &  24 \\
(11920) 1992 UY2 & 1 & SW low (noisy) & C & Phot. VIS & 24 \\
(12551) 1998 QQ39  & 1 &  -  &  V  &  Phot. VIS  &  24 \\
(12860) Turney  & 0 &   shorter band I centre  &  S  &  Phot. VIS  &  24, 24 \\
(13133) Jandecleir  & 1 &  SW medium  &  S  &  Phot. VIS  &  24 \\
(13704) Aletesi  & 0 &   shorter band I centre  &  C  &  Phot. VIS  &  24 \\
(13714) Stainbrook & 0 & noisy & S & Phot. VIS & 24 \\
(13743) Rivkin  & 0 &   shorter band I centre  &  V  &  Phot. VIS  &  24 \\
(13839) 1999 XF29  & 1 &  -  &  S  &  Phot. VIS  &  24 \\
(14108) 1998 OA13  & 0 &   shorter band I centre  &  -  &  -   &  - \\
(14489) 1994 UW  & 0 &  -  &  V  &  Phot. VIS  &  23 \\
(14511) Nickel  & 1 &  bump instead of band - SW low   &  -  &  -   &  - \\
(14562) 1997 YQ19 & 0 & noisy & V & Spec VISNIR & 25 \\
(15031) Lemus  & 0 &   shorter band I centre  &  V  &  Phot. NIR  &  17 \\
(15088) Licitra  & 1 &  SW low  &  S  &  Phot. VIS  &  24 \\
(15759) 1992 GM4  & 0 &   shorter band I centre  &  V  &  Phot. VIS  &  24 \\
(15989) 1998 XK39  & 1 &  -  &  V  &  Phot. VIS  &  24 \\
(16866) 1998 AR  & 0 &   no clear band I  &  S  &  Phot. VIS  &  24 \\
(16962) Elizawoolard  & 0 &  shorter band I centre  &  C  &  Phot. VIS  &  24 \\
(17225) Alanschorn  & 0 &   shorter band I centre  &  -  &  -   &  - \\
(17240) Gletorrence  & 1 &  -  &  S  &  Phot. VIS  &  24 \\
(17739) 1998 BY15  & 1 &  SW low  &  V  &  Phot. NIR  &  17 \\
(17821) Bolsche  & 1 &  lower quality spectrum - SW low   &  C  &  Phot. VIS  &  24 \\
(17882) Thielemann & 1 & SW low & V & Phot. VIS & 24 \\
(17904) Annekoupal  & 0 &  shorter band I centre  &  S  &  Phot. VIS  &  24 \\
(17943) 1999 JZ6 & 1 & SW low & V & Phot. VIS & 24 \\
(17951) Fenska  & 0 &  shorter band I centre  &  -  &  -   &  - \\
(18102) Angrilli  & 0 &  shorter band I centre  &  -  &  -   &  - \\
(18143) 2000 OK48 & 1 & SW medium & A & Phot. VIS & 10, 23, 24 \\
(18280) 4245 T-3  & 0 &  more similar to a V type  &  S  &  Phot. VIS  &  24 \\
(18344) 1989 TN11  & 1 &  SW low  &  V  &  Phot. VIS  &  24 \\
(19230) Sugazi  & 0 &   shorter band I centre  &  V  &  Phot. NIR  &  17 \\
(19281) 1996 AP3  & 0 &  -  &  V  &  Spec. VISNIR  &  18 \\
(19487) Rosscoleman  & 0 &  shorter band I centre   &  -  &  -  &  - \\
(19589) Kirkland & 0 & noisy & V & Phot. VIS & 24 \\
(19754) Paclements  & 1 &  SW high or medium   &  S  &  Phot. VIS  &  10, 23, 24 \\
(19978) 1989 TN6  & 1 &  SW low  &  V  &  Phot. VIS  &  10, 23 \\
(20079) 1994 EP  & 0 &   shorter band I centre  &  V  &  Phot. VIS  &  24 \\
(20157) 1996 TS18 & 0 & shorter band I centre & S & Phot. VIS & 24 \\
(20237) Clavius  & 0 &  no clear band I  &  -  &  -   &  - \\
(20289) Nettimi  & 1 &  SW low (noisy)  &  -  &  -  &  - \\
(20454) Pedrajo  & 1 &  noisy  &  S  &  Phot. VIS  &  24 \\
(20955) 2387 T-3  & 0 &  shorter band I centre  &  S  &  Phot. VIS  &  24 \\
(21318) 1996 XU26 & 1 & SW low & - & -  & - \\
(21435) Aharon  & 0 &  noisy  &  -  &  -   &  - \\
(21891) Andreabocelli  & 0 &  shorter band I centre  &  -  &  -   &  - \\
(22113) 2000 RH9  & 0 &   shorter band I centre  &  V  &  Phot. VIS  &  10, 23 \\
(22197) 3555 P-L  & 0 &  shorter band I centre  &  C  &  Phot. VIS  &  24 \\
(22322) Bodensee  & 0 &  shorter band I centre  &  V  &  Phot. VIS  &  24 \\
(23306) Adamfields  & 0 &  shorter band I centre  &  S  &  Phot. VIS  &  24 \\
(23502) 1992 DE3  & 0 &  shorter band I centre  &  -  &  -   &  - \\
(23595) 1995 VR11  & 0 &   shorter band I centre  &  C  &  Phot. VIS  &  24 \\
(23766) 1998 MZ23  & 1 &  SW low  &  S  &  Phot. VIS  &  24 \\
(24286) 1999 XU188  & 1 &  -  &  S  &  Phot. VIS  &  24 \\
(24569) 9609 P-L  & 1 &   SW low or medium  &  S  &  Phot. NIR  &  17 \\
(24684) 1990 EU4 & 1 & SW low & S & Phot. NIR & 17 \\
(24892) 1997 AD3  & 1 & -  &  -  &  -   &  - \\
(25434) Westonia  & 0 &  shorter band I centre   &  V  &  Phot. VIS  &  23, 24 \\
(25752) 2000 BE8 & 0 & noisy + bad BP-RP alignment & - & -  & - \\
(25808) 2000 CK103 & 0 & flatter red part & S & Phot. VIS & 24 \\
(26084) 1981 EK17  & 1 &  SW low  &  S  &  Phot. VIS  &  24 \\
(26417) Michaelgord  & 0 &  bad BP-RP overlapping   &  V  &  Phot. VIS  &  10, 23, 24 \\
(26573) 2000 EG87  & 1 &  -  &  V  &  Phot. VIS  &  24 \\
(26851) Sarapul  & 1 &  SW low  &  -  &  -   &  - \\
(27106) Jongoldman & 0 & shorter band I centre & V & Phot. VIS & 24 \\
(27162) 1999 AM6  & 0 &  shorter band I centre  &  S  &  Phot. VIS  &  24 \\
(27262) 1999 XT184 & 1 & bad RP & X & Phot. VIS & 24 \\
(27390) Kyledavis & 0 & shorter band I centre & - & -  & - \\
(27399) Gehring  & 0 &   shorter band I centre  &  C  &  Phot. VIS  &  24 \\
(27876) 1996 BM4 & 1 & SW low & S & Phot. VIS & 24 \\
(27884) 1996 EZ1  & 1 &  SW low  &  S  &  Phot. VIS  &  24 \\
(28132) Karenzobel & 1 & SW low & S & Phot. VIS & 24 \\
(28162) 1998 VD14  & 1 &  -  &  -  &  -   &  - \\
(28291) 1999 CX52  & 0 &   shorter band I centre  &  V  &  Spec. VIS  &  9 \\
(29171) 1990 QK3  & 1 &  bump instead of 0.65~\SI{}{\micro\meter} band - SW low  &  -  &  -   &  - \\
(29269) 1993 FD25  & 0 &   shorter band I centre  &  C  &  Phot. VIS  &  24 \\
(30426) Philtalbot  & 1 &  SW low   &  V  &  Phot. VIS  &  23 \\
(30751) 1981 EL29  & 0 &   shorter band I centre  &  S  &  Phot. VIS  &  24 \\
(30769) 1984 ST2  & 1 &  -   &  -  &  -   &  - \\
(30781) 1988 CR2 & 0 & shorter band I centre & C & Phot. VIS & 24 \\
(30820) 1990 RU2  & 0 &  more similar to a V type  &  S  &  Phot. VIS  &  24 \\
(30834) 1990 VR6  & 1 &  SW low  &  V  &  Phot. VIS  &  10, 23 \\
(30892) 1993 FR18  & 0 &   shorter band I centre  &  A  &  Phot. VIS  &  23 \\
(31060) 1996 TB6  & 1 &  SW medium   &  SQ  &  Phot. VIS  &  7 \\
(31414) Rotarysusa & 0 & shorter band I centre & V & Spec. VISNIR & 25 \\
(31544) 1999 DZ5  & 0 &  shorter band I centre  &  V  &  Phot. VIS  &  24 \\
(31572) 1999 FM22 & 0 & shorter band I centre & V & Phot. VIS & 24 \\
(31622) 1999 GL19 & 0 & shorter band I centre & - & -  & - \\
(32168) 2000 NP9  & 0 &  shorter band I centre  &  -  &  -   &  - \\
(32449) Crystalmiller  & 0 &  shorter band I centre   &  S  &  Phot. VIS  &  24 \\
(32590) Cynthiachen  & 0 &  shorter band I centre SW low  &  V  &  Phot. VIS  &  10, 23 \\
(33418) Jacksonweaver & 1 & - & V & Phot. VIS & 10, 23 \\
(33562) Amydunphy & 0 & different red part & V & Phot. NIR & 17 \\
(33881) 2000 JK66  & 0 &  -   &  V  &  Spec. VISNIR  &  20 \\
(33947) 2000 ML1  & 1 &  SW low   &  S  &  Phot. VIS  &  24 \\
(34698) 2001 OD22  & 0 &   shorter band I centre  &  V  &  Spec. NIR  &  16 \\
(34706) 2001 OP83  & 0 &  Vesta family  &  V  &  Spec. NIR  &  14 \\
(35193) 1994 CG14  & 0 &   no clear band I  &  C  &  Phot. VIS  &  24 \\
(35364) Donaldpray  & 1 &  SW low  &  V  &  Phot. VIS  &  10 \\
(36360) 2000 OH3  & 0 &   shorter band I centre  &  S  &  Phot. VIS  &  24 \\
(36363) 2000 OB5  & 0 &  shorter band I centre  &  S  &  Phot. VIS  &  24 \\
(36431) 2000 PJ12  & 1 &  -   &  V  &  Phot. VIS  &  7 \\
(36798) 2000 SA43  & 0 &  shorter band I centre + noisy  &  S  &  Phot. VIS  &  24 \\
(37306) 2001 KW46  & 0 &  no clear band I  &  -  &  -   &  - \\
(37386) 2001 WG29 & 0 & shorter band I centre & V & Phot. NIR & 17 \\
(39940) 1998 FR99  & 1 &  SW low (bad BP)  &  -  &  -   &  - \\
(40056) 1998 KT44  & 0 &   shorter band I centre  &  C  &  Phot. VIS  &  24 \\
(41574) 2000 SQ1  & 0 &  no clear band I   &  -  &  -   &  - \\
(41765) 2000 VV35  & 0 &   shorter band I centre  &  X  &  Phot. VIS  &  24 \\
(41894) 2000 WH121  & 1 &  SW low   &  -  &  -   &  - \\
(42644) 1998 FE67 & 0 & bump instead of 0.65~\SI{}{\micro\meter} band & V & Phot. NIR & 17 \\
(42822) 1999 NT13 & 1 & SW medium & S & Phot. VIS & 24, 24 \\
(43278) 2000 ES109  & 1 &  SW low   &  C  &  Phot. VIS  &  24 \\
(43302) 2000 GE114 & 0 & shorter band I centre & V & Phot. VIS & 24 \\
(43388) 2000 WA61  & 0 &  shorter band I centre  &  V  &  Phot. NIR  &  17 \\
(44150) 1998 HC108  & 1 &  -   &  V  &  Phot. VIS  &  10, 23 \\
(44162) 1998 HC148  & 1 &  SW low   &  C  &  Phot. VIS  &  24 \\
(44322) 1998 RZ42  & 1 &  SW medium   &  S  &  Phot. VIS  &  24 \\
(44711) Carp  & 0 &  no clear band I  &  S  &  Phot. VIS  &  24 \\
(44940) 1999 VH53  & 0 &  shorter band I centre   &  C  &  Phot. VIS  &  24, 24 \\
(45417) 2000 AZ151  & 0 &  shorter band I centre  &  -  &  -   &  - \\
(45787) 2000 OJ24  & 1 &  SW low   &  -  &  -   &  - \\
(46701) Interrante & 0 & shorter band I centre & V & Phot. VIS & 23 \\
(47232) 1999 VQ36 & 1 & good agreement between 500 and 950 nm & C & Phot. VIS & 24 \\
(47398) 1999 XC116  & 0 &  bump instead of 0.65~\SI{}{\micro\meter} band   &  V  &  Phot. VIS  &  23 \\
(47463) 1999 XE258  & 0 &  shorter band I centre   &  -  &  -   &  - \\
(48039) 2001 DT69  & 1 &  SW low  &  V  &  Phot. VIS  &  23 \\
(48114) 2001 FW77 & 0 & different blue part & S & Phot. VIS & 10, 23 \\
(48323) 2002 NN33  & 0 &  low quality spectrum  &  S  &  Phot. VIS  &  24 \\
(48632) 1995 SV29  & 0 &  more similar to a V type  &  V  &  Phot. VIS  &  10 \\
(49101) 1998 RE76  & 1 &  -   &  V  &  Phot. VIS  &  10, 23 \\
(49141) 1998 SM41  & 1 &  SW medium (or A type?)  &  A  &  Phot. VIS  &  10, 23 \\
(49901) 1999 XK164  & 0 &  shorter band I centre   &  S  &  Phot. VIS  &  24 \\
(50139) 2000 AH129  & 0 &  no clear band I  &  -  &  -   &  - \\
(50236) 2000 BB3  & 0 &  shorter band I centre SW low  &  V  &  Phot. VIS  &  24 \\
(51379) 2001 BY7 & 1 & SW medium (noisy) & C & Phot. VIS & 24 \\
(51443) 2001 FN27  & 0 &  bump instead of band   &  V  &  Phot. NIR  &  17 \\
(51659) Robohachi  & 1 &  SW low (noisy)  &  S  &  Phot. VIS  &  24 \\
(52216) 5014 T-3 & 0 & shorter band I centre & V & Phot. VIS & 24 \\
(52408) 1993 TJ34  & 1 &  SW medium   &  -  &  -   &  - \\
(52995) 1998 UJ32  & 0 &  shorter band I centre  &  V  &  Phot. NIR  &  17 \\
(53417) 1999 NP38  & 1 &  SW low   &  -  &  -   &  - \\
(53425) 1999 SO4  & 0 &  noisy  &  S  &  Phot. VIS  &  10, 23 \\
(53593) 2000 CJ58  & 0 &  shorter band I centre  &  S  &  Phot. VIS  &  24 \\
(53661) 2000 DU62  & 1 &  SW low   &  S  &  Phot. VIS  &  24 \\
(53899) 2000 FM49  & 1 &  SW low or medium   &  -  &  -   &  - \\
(54061) 2000 GX134  & 0 &  shorter band I centre  &  -  &  -   &  - \\
(55549) 2001 XC59 & 1 & noisy but plausible & S & Phot. VIS & 24 \\
(55831) 1995 XL & 0 & bad BP-RP alignment & S & Phot. NIR & 17 \\
(56348) 2000 AH69  & 0 &  shorter band I centre  &  C  &  Phot. VIS  &  24 \\
(56561) Jaimenomen & 1 & SW low & - & - & - \\
(56585) 2000 JZ29  & 0 &  shorter band I centre  &  Q  &  Phot. VIS  &  24 \\
(56696) 2000 LQ26  & 0 &  shorter band I centre  &  V  &  Phot. VIS  &  10, 23 \\
(56904) 2000 QP171 & 1 & - & C & Phot. VIS & 24 \\
(57857) 2001 XJ203 & 0 & shorter band I centre & - & -  & - \\
(58640) 1997 WH18  & 1 &  SW low  &  -  &  -   &  - \\
(59228) 1999 CH & 0 & shorter band I centre & V & Phot. VIS & 10, 23 \\
(59530) 1999 JU24  & 0 &  shorter band I centre  &  -  &  -   &  - \\
(59686) 1999 JS108  & 0 &  shorter band I centre  &  -  &  -   &  - \\
(60285) 1999 XR106  & 0 &  shorter band I centre  &  S  &  Phot. VIS  &  24 \\
(60584) 2000 EW132  & 0 &  shorter band I centre  &  S  &  Phot. VIS  &  24 \\
(61098) 2000 LY28 & 1 & SW low & V & Phot. VIS & 24 \\
(61203) 2000 OY4  & 0 &  -   &  V  &  Phot. VIS  &  24 \\
(61682) 2000 QV124  & 0 &   shorter band I centre  &  C  &  Phot. VIS  &  24 \\
(63366) 2001 HK4 & 0 & different red part & V & Phot. VIS & 10 \\
(63438) 2001 MY28  & 0 &   no clear band I  &  -  &  -   &  - \\
(64252) 2001 TL168  & 0 &   shorter band I centre  &  A  &  Phot. VIS  &  24 \\
(64458) 2001 VF35  & 1 &  SW low  &  V  &  Phot. NIR  &  17 \\
(64948) 2001 YH124 & 0 & noisy & S & Phot. VIS & 24 \\
(65707) 1992 PY1  & 0 &  bad quality spectrum   &  -  &  -   &  - \\
(66679) 1999 TD29  & 0 &   shorter band I centre  &  V  &  Phot. VIS  &  24 \\
(68765) 2002 EE99 & 0 & shorter band I centre & - &  - & - \\
(69595) 1998 FK11  & 0 &   shorter band I centre  &  V  &  Phot. VIS  &  24 \\
(69628) 1998 FD62  & 0 &   shorter band I centre  &  S  &  Phot. VIS  &  24 \\
(74107) 1998 QM37  & 1 &  SW low? Bad BP-RP overlapping?  &  -  &  -   &  - \\
(75323) 1999 XY47  & 1 &  SW low  &  -  &  -   &  - \\
(75441) 1999 XB129  & 0 &   shorter band I centre  &  S  &  Phot. VIS  &  24 \\
(77584) 2001 KP14 & 0 & noisy & S & Phot. VIS & 24 \\
(77590) 2001 KM17  & 0 &   shorter band I centre  &  V  &  Phot. NIR  &  17 \\
(78034) 2002 JF82  & 0 &   shorter band I centre  &  V  &  Phot. VIS  &  7 \\
(79137) 1991 PD15 & 0 & no band I & - & -  & - \\
(80356) 1999 XM124  & 0 &   no clear band I  &  Ad  &  Phot. NIR  &  17 \\
(80863) 2000 DT27  & 0 &  more similar to a V type  &  V  &  Phot. VIS  &  10 \\
(85301) 1994 UM5 & 0 & shorter band I centre & - & -  & - \\
(87093) 2000 LW6  & 1 &  -   &  V  &  Phot. VIS  &  23 \\
(87216) 2000 OG38  & 1 &  SW low (bad BP spectrum)  &  -  &  -   &  - \\
(88912) 2001 TS8  & 0 &   no clear band I  &  V  &  Phot. VIS  &  24 \\
(88955) 2001 TW42  & 1 &  -   &  S  &  Phot. VIS  &  24 \\
(89776) 2002 AL90  & 1 &  SW low  &  -  &  -   &  - \\
(89952) 2002 JB20  & 1 &  SW low   &  S  &  Phot. VIS  &  24 \\
(90604) 4813 P-L  & 1 &   bad red part  &  S  &  Phot. VIS  &  24 \\
(90843) 1995 YZ22  & 1 &  SW medium  &  -  &  -   &  - \\
(90855) 1996 GZ8 & 0 & bump instead of band & C & Phot. VIS & 24 \\
(91343) 1999 JP30  & 0 &  shorter band I centre  &  V  &  Phot. NIR  &  17 \\
(92593) 2000 PN16  & 1 &  SW low  &  -  &  -   &  - \\
(98482) 2000 UL101  & 0 &  noisy   &  S  &  Phot. VIS  &  24 \\
(98745) 2000 YB47  & 0 &   shorter band I centre  &  V  &  Phot. NIR  &  17 \\
(99714) 2002 JQ41  & 1 &  SW medium  &  S  &  Phot. VIS  &  24 \\
(102071) 1999 RK139  & 0 &  shorter band I centre  &  V  &  Phot. VIS  &  10, 23 \\
(102107) 1999 RL164  & 0 &  shorter band I centre  &  V  &  Phot. VIS  &  10 \\
(102195) 1999 ST10  & 0 &  noisy  &  -  &  -   &  - \\
(102469) 1999 TC237  & 0 &  bap BP RP overlapping?   &  V  &  Phot. VIS  &  23 \\
(108139) 2001 GL11  & 1 &  SW low   &  V  &  Phot. VIS  &  7 \\
(108199) 2001 HX21  & 0 &   no clear band I  &  -  &  -   &  - \\
(112326) 2002 MM4  & 1 &  SW low  &  V  &  Phot. VIS  &  23 \\
(114486) 2003 AJ57  & 0 &  -   &  -  &  -   &  - \\
(119385) 2001 TU7  & 0 &  bump instead of 0.65~\SI{}{\micro\meter} band, bad blue and red parts   &  V  &  Phot. NIR  &  17 \\
(122122) 2000 JM16  & 1 &  SW low  &  V  &  Phot. VIS  &  23 \\
(122125) 2000 JO17  & 1 &  SW medium  &  S  &  Phot. VIS  &  10, 23 \\
(125002) 2001 TJ154 & 0 & shorter band I centre & - &  - & - \\
(127422) 2002 OX11  & 0 &  low quality spectrum   &  S  &  Phot. VIS  &  10, 23 \\
(128450) 2004 NX24  & 1 &  SW low  &  -  &  -   &  - \\
(130988) 2000 WT141  & 0 &  NEA  &  V  &  Spec. VIS  &  13 \\
(133245) 2003 RL2  & 0 &   shorter band I centre  &  -  &  -   &  - \\
(134693) 1999 XP67 & 0 & noisy & - & -  & - \\
(134916) 2000 YP53  & 1 &  bad RP spectrum, SW low  &  -  &  -   &  - \\
(149372) 2002 XC71  & 0 &  bad agreement before 700 nm   &  -  &  -   &  - \\
(150544) 2000 SG164  & 0 &   noisy  &  X  &  Phot. VIS  &  23 \\
(158242) 2001 TM24 & 0 & bad BP-RP alignment & V & Phot. VIS & 23 \\
(163804) 2003 QQ88  & 0 &  noisy  &  S  &  Phot. VIS  &  7 \\
(179587) 2002 LS2  & 0 &  -   &  S  &  Phot. VIS  &  15 \\
(180757) 2004 NE33  & 0 &  -   &  -  &  -   &  - \\
(190138) 2005 RW27  & 0 &   shorter band I centre  &  -  &  -   &  - \\
(190664) 2000 YX90  & 0 &   bad BP-RP overlapping  &  -  &  -   &  - \\
(205560) 2001 SC282  & 1 &  noisy but plausible  &  -  &  -   &  - \\
(230762) 2003 WP192  & 1 &  SW medium   &  -  &  -   &  - \\
(310436) 2000 AB169 & 1 & noisy but plausible & - & -  & - \\
\hline  
\end{longtable}
\end{ThreePartTable}
} 

\begin{ThreePartTable}
\setlength\tabcolsep{3.5pt} 
\footnotesize\setlength{\tabcolsep}{2.6pt}
    \sisetup{table-format=-1.4}
\begin{TableNotes}
  \item[\bfseries Note: ] The information in the table are the number and name of the 58 asteroids, if they are accepted or not as a good match for EC~002 (1 if accepted, 0 if not), some notes about the visual inspection, the spectral type of the asteroid if determined and the method and relevant references associated (Ref. column). The taxonomic scheme used for the type of each asteroid is the one used in the reference papers associated. Spec. stands for Spectroscopy and Phot. for Photometry. 
\end{TableNotes}
\begin{longtable}{l c c c c c}
\caption{Asteroids found as a match to the powder and raw slab samples of EC~002 with a curve-matching method. \label{tab:slabPowderCM}} \\ 
\hline 
Asteroid & Acceptance & Notes & Type & Method & Ref. \\ 
\hline 
\endfirsthead 
{\tablename\ \thetable\ -- continued.} \\ 
\hline 
\endhead 
\hline 
\endfoot
\hline
\insertTableNotes
\endlastfoot
(6853) Silvanomassaglia & 1 & - & V & Phot. NIR & 17 \\
(10156) 1994 VQ7 & 1 & - & V & Phot. VIS & 24 \\
(13743) Rivkin & 0 &  shorter band I centre & V & Phot. VIS & 24 \\
(16856) Banach & 1 & - & S & Phot. VIS & 24 \\
(17056) Boschetti & 1 & - & S & Phot. VIS & 24 \\
(20289) Nettimi & 0 &  noisy and unclear band I & - & -  & - \\
(20454) Pedrajo & 1 & - & S & Phot. VIS & 24 \\
(23522) 1992 WC9 & 0 &  shorter band I centre & V & Phot. NIR & 17 \\
(24143) 1999 VY124 & 0 &  noisy & C & Phot. VIS & 24 \\
(24892) 1997 AD3 & 1 & - & - &  - & - \\
(26399) Rileyennis & 0 &  shorter band I centre & - & -  & - \\
(26420) 1999 XL103 & 0 &  shorter band I centre & V & Phot. VIS & 23 \\
(27106) Jongoldman & 0 &  shorter band I centre & V & Phot. VIS & 24 \\
(27627) 2038 P-L & 0 &  shorter band I centre & V & Phot. VIS & 24 \\
(30000) Camenzind & 0 &  shallow slope & V & Phot. VIS & 10, 23, 24 \\
(30081) Zarinrahman & 0 &  shorter band I centre & S & Phot. VIS & 24 \\
(38690) 2000 QS29 & 0 &  unclear band I & S & Phot. VIS & 24 \\
(40693) 1999 RX229 & 0 &  unclear band I & C & Phot. VIS & 24 \\
(44691) 1999 RF221 & 0 &  shallow slope & C & Phot. VIS & 24 \\
(47327) 1999 XZ25 & 0 &  shorter band I centre & V & Phot. VIS & 10, 23 \\
(48632) 1995 SV29 & 0 &  shorter band I centre & V & Phot. VIS & 10 \\
(50488) 2000 DA86 & 0 &  shallow slope & - &  - & - \\
(51659) Robohachi & 0 &  noisy & S & Phot. VIS & 24 \\
(51688) 2001 KW12 & 0 &  shorter band I centre & S & Phot. VIS & 24 \\
(53561) 2000 CM22 & 0 &  noisy and unclear band I & S & Phot. VIS & 24 \\
(54062) 2000 GX135 & 1 &  noisy but plausible & C & Phot. VIS & 24 \\
(55549) 2001 XC59 & 1 &   & S & Phot. VIS & 24 \\
(55866) 1997 PV4 & 0 &  shallow slope & V & Phot. VIS & 24 \\
(59686) 1999 JS108 & 0 &  shorter band I centre & - & -  & - \\
(61169) 2000 NY20 & 0 &  band red and blue parts & X & Phot. VIS & 24 \\
(63653) 2001 QQ109 & 1 & - & - & -  & - \\
(64181) 2001 TS64 & 0 &  shorter band I centre & V & Phot. VIS & 23 \\
(68814) 2002 GP66 & 0 &  shallow slope & - & -  & - \\
(77147) 2001 EV6 & 1 &  bad two last points & S & Phot. VIS & 24 \\
(77590) 2001 KM17 & 0 &  shorter band I centre & V & Phot. NIR & 17 \\
(77935) 2002 GM54 & 1 &  noisy but plausible & V & Phot. VIS & 24 \\
(78034) 2002 JF82 & 0 &  shorter band I centre & V & Phot. VIS & 7 \\
(80924) 2000 DJ73 & 0 &  noisy and unclear band I & C & Phot. VIS & 24 \\
(81448) 2000 GV123 & 0 &  shallow slope & S & Phot. VIS & 24 \\
(87010) 2000 JR55 & 0 &  shallow slope & C & Phot. VIS & 24, 24 \\
(88955) 2001 TW42 & 1 & - & S & Phot. VIS & 24 \\
(89556) 2001 XS98 & 1 &  except for last points & - & -  & - \\
(93893) 2000 WL141 & 0 &  unclear band I & S & Phot. VIS & 10, 23, 24 \\
(96353) 1997 VF3 & 0 &  flatter spectrum & C & Phot. VIS & 24 \\
(99722) 2002 JW46 & 0 &  flatter spectrum & S & Phot. VIS & 24 \\
(102107) 1999 RL164 & 0 &  shorter band I centre & V & Phot. VIS & 10 \\
(103308) 2000 AH55 & 0 &  unclear band I & - & -  & - \\
(119144) 2001 PH32 & 0 &  unclear band I & V & Phot. VIS & 10, 23 \\
(123113) 2000 SH361 & 1 & - & V & Phot. VIS & 23 \\
(124884) 2001 TE41 & 1 & -   & V & Phot. VIS & 10, 23 \\
(130988) 2000 WT141 & 0 &  NEA & V & Spec. VIS & 13 \\
(147124) 2002 TH129 & 0 &  less pronounced band & - & -  & - \\
(149372) 2002 XC71 & 0 &  bad agreement before 700 nm & - & -  & - \\
(153408) 2001 QV137 & 0 &  shorter band I centre & - & -  & - \\
(164121) 2003 YT1 & 1 &  RP noisy but plausible & V & Spec. VISNIR & 12 \\ 
(194248) 2001 TA199 & 0 &  flatter spectrum & - & -  & - \\
(205560) 2001 SC282 & 1 &  noisy but plausible & - & -  & - \\
(310436) 2000 AB169 & 1 &  noisy but plausible & - & -  & - \\
\hline
\end{longtable} 
\end{ThreePartTable}

\begin{table}
\begin{ThreePartTable}
\centering
\caption{Accepted asteroids as candidate matches to the three space-weathered modelled samples of EC~002. \label{tab:SWCM}}
\begin{tabular}{l c c c}
		\hline
		Asteroid & Type & Method & Ref \\ 
		\hline
        \textbf{SW low} \\ 
        \hline 
        (10131) Stanga  &  S  &  Phot. VIS  &  24 \\
        (15623) 2000 HU30  &  S  &  Phot. NIR  &  17 \\
        (18780) Kuncham  &  S  &  Phot. VIS  &  24 \\
        (20535) Marshburrows  &  L  &  Phot. VIS  &  24 \\
        (22276) Belkin  &  S  &  Phot. VIS  &  24 \\
        (22538) Lucasmoller  &  S  &  Phot. VIS  &  24 \\
        (24684) 1990 EU4  &  S  &  Phot. NIR  &  17 \\
        (27876) 1996 BM4  &  S  &  Phot. VIS  &  24 \\
        (32835) 1992 EO5  &  V  &  Phot. VIS  &  24 \\
        (33423) 1999 DK  & A  &  Phot. VIS  &  23 \\
        (33852) Baschnagel  &  V  &  Phot. VIS  &  24 \\
        (33934) 2000 LA30  &  S  &  Phot. VIS  &  24 \\
        (33947) 2000 ML1  &  S  &  Phot. VIS  &  24 \\
        (43278) 2000 ES109  &  C  &  Phot. VIS  &  24 \\
        (56561) Jaimenomen  &  - & - & -  \\
        (65504) 3544 P-L  &  V  &  Phot. NIR  &  17 \\
        (74378) 1998 XH11  &  S  &  Phot. NIR  &  17 \\
        (79827) 1998 WU3  &  -  &  -   & -  \\
        (89952) 2002 JB20  &  S  &  Phot. VIS  &  24 \\
        (100440) 1996 PJ6  &  -  &  -   & -  \\
        (103308) 2000 AH55  & -   & - & -  \\
        (108139) 2001 GL11  &  V  &  Phot. VIS  &  7 \\
        (112326) 2002 MM4  &  V  &  Phot. VIS  &  23 \\
        \hline 
        \textbf{SW medium} \\ 
        \hline
        (42822) 1999 NT13  &  S  &  Phot. VIS  &  24 \\
        (44322) 1998 RZ42  &  S  &  Phot. VIS  &  24 \\
        (68089) 2000 YS108  &  -  & - & -  \\
        (68946) 2002 PX138  &  S  &  Phot. VIS  &  24 \\
        (93797) 2000 WO43  &  S  &  Phot. VIS  &  10 \\
        (108899) 2001 PP5  &  -  &  -   & -  \\
        (145532) 2006 FD42  &  -  &   - &  - \\
        (230762) 2003 WP192  &  -  & -   & -  \\
        \hline 
        \textbf{SW high} \\ 
        \hline
        (33809) 1999 XK152  &  C  &  Phot. VIS  &  24 \\
        \hline
\end{tabular}
\begin{tablenotes}
      \small
      \item[\bfseries Note: ] This selection has been done after visual inspection of 269 asteroids for the low space-weathered sample, 223 asteroids for the medium space weathering and 12 asteroids for the high space weathering. The references associated with the numbers in the Ref. column are given in appendix. The taxonomic scheme used for the type of each asteroid is the one used in the reference papers associated. SW stands for space weathering.
    \end{tablenotes}
  \end{ThreePartTable}
\end{table}

\textbf{Note:} The references are (1) \cite{xu1995}, (2) \cite{bus2002tax}, (3) \cite{lazzaro2004}, (4) \cite{AlvarezCandal2006}, (5) \cite{demeo2009},  (6) \cite{Moskovitz2010}, (7) \cite{carvano2010}, (8) \cite{DeSanctis2011}, (9) \cite{solontoi2012}, (10) \cite{demeo2013}, (11) \cite{Jasmim2013}, (12) \cite{Sanchez2013}, (13) \cite{Riberio2014}, (14) \cite{Lindsay2015}, (15) \cite{Carry2016}, (16) \cite{Hardersen_2018}, (17) \cite{Popescu_2018_classification}, (18)\cite{Medeiros2019}, (19) \cite{DeMeo2019}, (20) \cite{Binzel2019}, (21) \cite{Matlovic2020}, (22) \cite{Migliorini2021}, (23) \cite{SergeyevCarry2021}, (24) \cite{Sergeyev2022}, (25) \cite{Mahlke2022}

\section{Spectra of the asteroids matching the spectra of EC~002}
\begin{figure*}[t]
  \centering
  \begin{adjustbox}{clip,trim=0cm 0.04cm 0cm 0cm,max width=\textwidth}
  \input{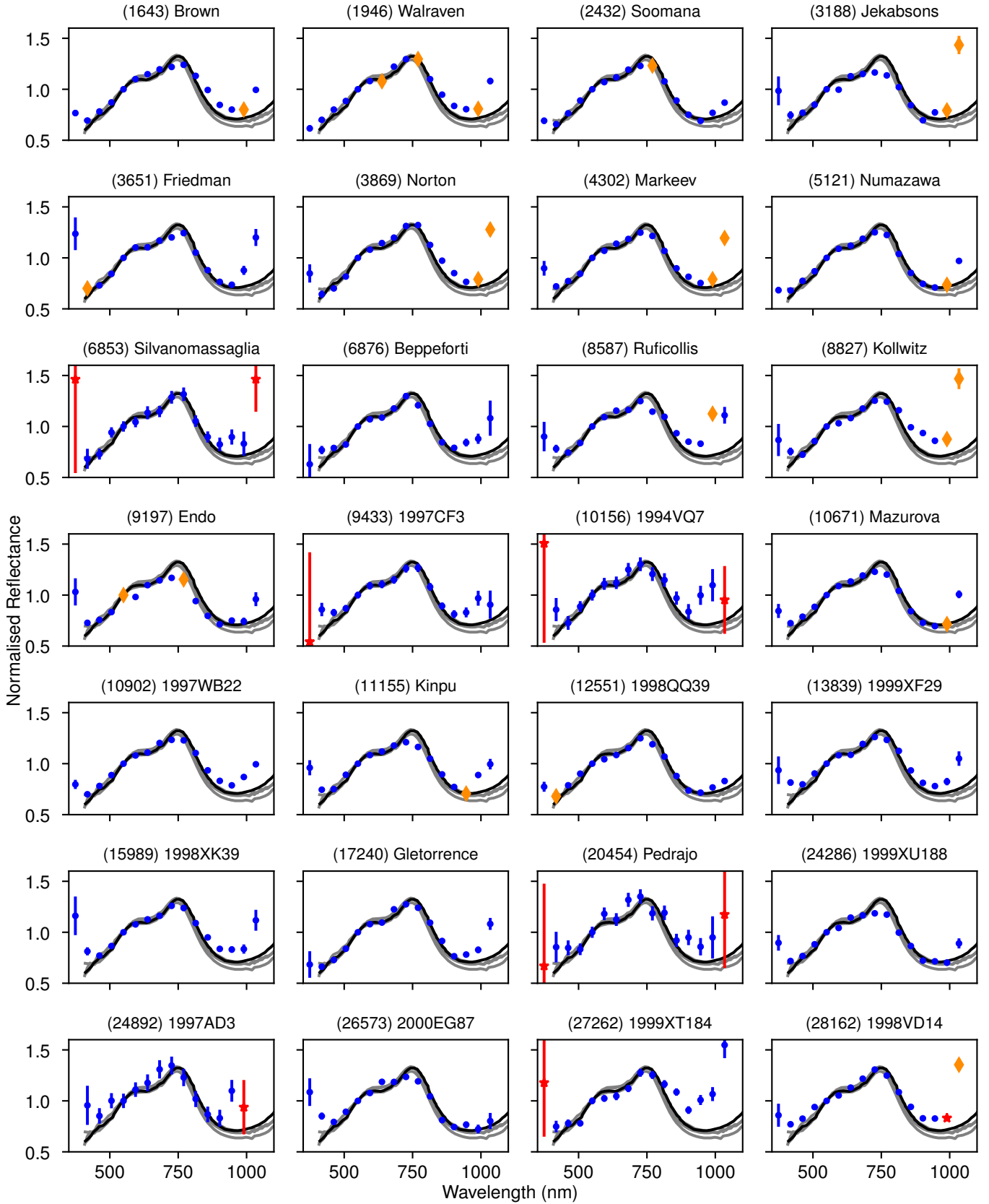}
  \end{adjustbox}
    \caption{Spectra of the 41 asteroids found in the 'possible matches area', validated as matches of EC~002 after visual inspection. The spectra are normalised at 550~nm. Black continuous line: spectrum of the powder sample of the meteorite, grey lines: spectra of the raw slab samples. The 16 bands of the Gaia asteroid spectra are given a colour and a symbol according to the value of the flag associated to the band: blue circle if flag=0, orange diamond if flag=1 and red star if flag=2. This way of showing the asteroid spectra applies for every figure hereafter.}
    \label{fig:matchesnoSW}
\end{figure*}

\renewcommand{\thefigure}{B.1}
\begin{figure*}[t]
  \centering
    \input{Img/matches_noSW_test2.pgf}
    \caption{continued.}
\end{figure*}
%

%
\renewcommand{\thefigure}{B.2}
\begin{figure*}[t]
  \centering
  \begin{adjustbox}{clip,trim=0.04cm 0cm 0.04cm 0cm,max width=\textwidth}
  \input{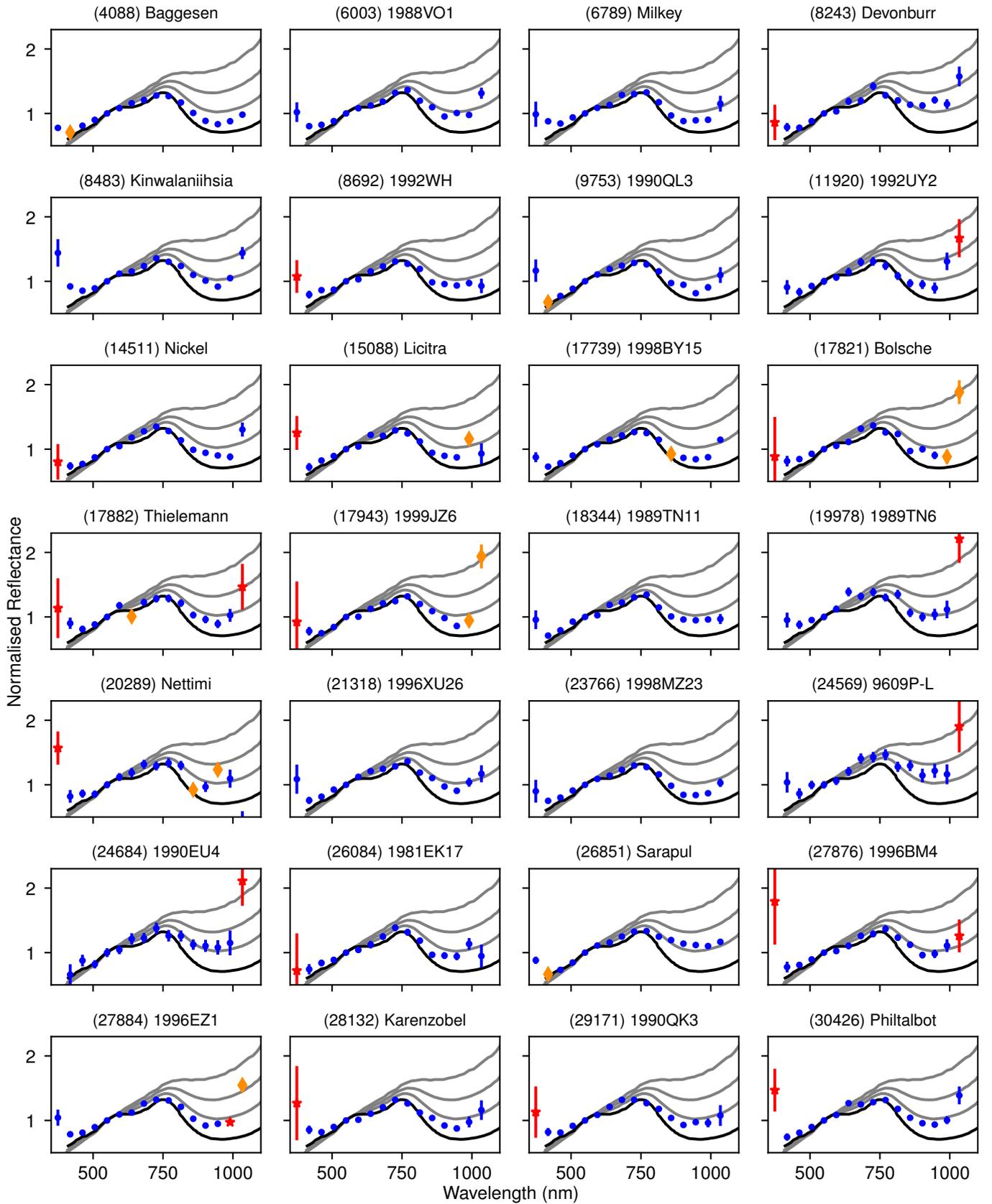}
  \end{adjustbox}
    \caption{ Same as Fig.~\ref{fig:matchesnoSW} but with the 56 asteroids visually validated as matches of the low space-weathered EC~002. The space-weathered spectra of EC~002 are shown in grey lines.}
    \label{fig:matchesSWlow}
\end{figure*}

\renewcommand{\thefigure}{B.2}
\begin{figure*}[t]
  \centering
    \input{Img/matches_SWarea_SWlow2.pgf}
    \caption{continued.}
\end{figure*}
%

%
\renewcommand{\thefigure}{B.3}
\begin{figure*}[t]
  \centering
  \input{Img/matches_SWarea_SWmed_v2.pgf}
    \caption{ Same as Fig.~\ref{fig:matchesSWlow} but with the 12 asteroids visually validated as matches of the medium space-weathered EC~002. }
    \label{fig:matchesSWmed}
\end{figure*}
%

%
\renewcommand{\thefigure}{B.4}
\begin{figure*}[t]
  \centering
  \input{Img/matches_SWarea_SWhigh_v2.pgf}
    \caption{ Same as Fig.~\ref{fig:matchesSWlow} but with the two asteroids visually validated as matches of the high space-weathered EC~002. }
    \label{fig:matchesSWhigh}
\end{figure*}

\renewcommand{\thefigure}{B.5}
\begin{figure*}[t]
  \centering
  \input{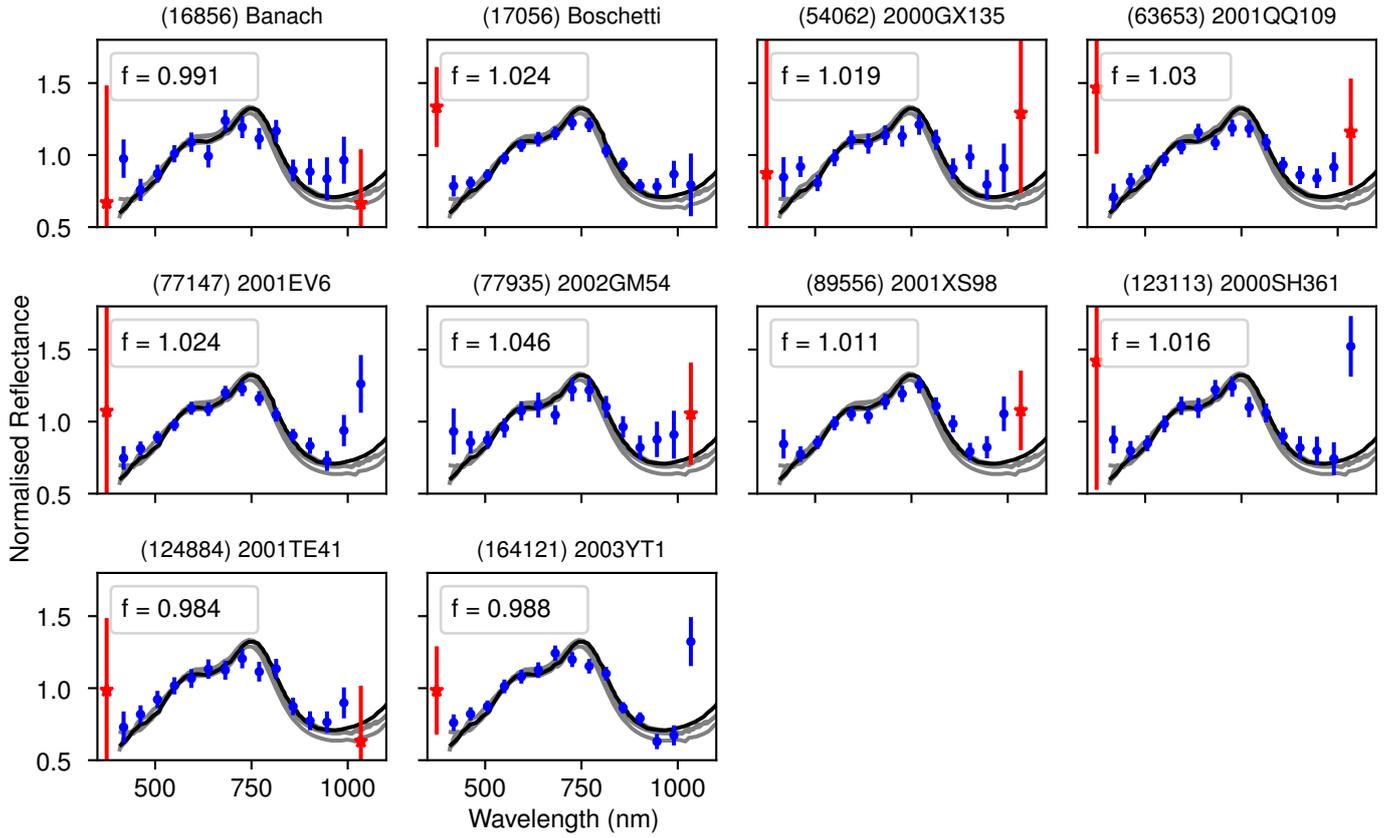}
    \caption{Spectra of the ten asteroids found with the curve matching method only, validated as matches of EC~002 after visual inspection. The spectra are normalised with a scaling factor f, here the meteorite spectrum was divided by the scaling factor. The spectra of the powder sample of the meteorite is shown in black continuous line, and the raw slab samples spectra are shown in grey lines. As previously, the 16 bands of the Gaia asteroid spectra are shown with a colour and a symbol associated to their flag number.}
    \label{fig:matchesnoSW_CM}
\end{figure*}

\renewcommand{\thefigure}{B.6}
\begin{figure*}[t]
  \centering
  \input{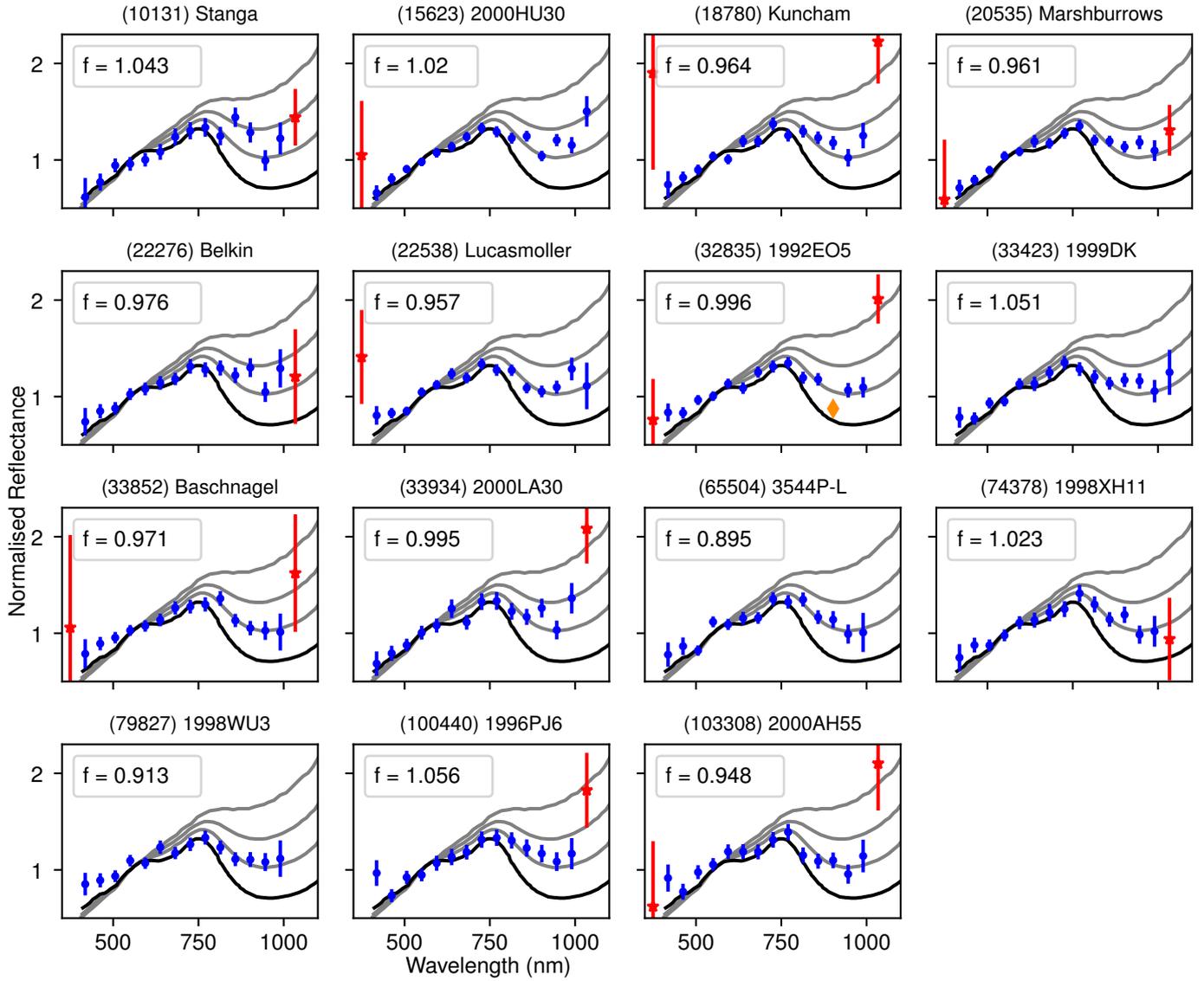}
    \caption{ Same as Fig.~\ref{fig:matchesnoSW_CM} but with the 15 asteroids visually validated as matches of the low space-weathered EC~002. Here the spectra of the powder sample of the meteorite is shown in black continuous line, and the space-weathered spectra are shown in grey lines.}
    \label{fig:matchesSWlow_CM}
\end{figure*}
\renewcommand{\thefigure}{B.7}
\begin{figure*}[t]
  \centering
  \input{Img/matches_CM_SWmed.pgf}
    \caption{ Same as Fig.~\ref{fig:matchesSWlow_CM} but with the 8 asteroids visually validated as matches of the medium space-weathered EC~002. }
    \label{fig:matchesSWmed_CM}
\end{figure*}
\renewcommand{\thefigure}{B.8}
\begin{figure*}[t]
  \centering
  \input{Img/matches_CM_SWhigh.pgf}
    \caption{ Same as Fig.~\ref{fig:matchesSWlow_CM} but with the asteroid visually validated as match of the high space-weathered EC~002. }
    \label{fig:matchesSWhigh_CM}
\end{figure*}
\end{appendix}

\end{document}